\def\fig{\frenchspacing Fig.~}
\documentclass[usenatbib]{mn2e} 
\usepackage{amsmath}
\usepackage{epsfig}
\usepackage{amssymb}

\title[A search for concentric rings with unusual variance]{A search
  for concentric rings with unusual variance in the 7-year WMAP
  temperature maps using a fast convolution approach}

\author[P.~Bielewicz, B.~D.~Wandelt, A.~J.~Banday]{P.~Bielewicz$^{1,2,3}$ \thanks{E-mail:
    pbielew@fuw.edu.pl}, B.~D.~Wandelt$^{4,5}$, A.~J.~Banday$^{1,2}$,\\
  $^1$ Universit\'e de Toulouse, UPS-OMP, IRAP, Toulouse, France \\ $^2$ CNRS, IRAP, 9 Av. colonel
  Roche, BP 44346, F 31028 Toulouse cedex 4, France \\  
$^3$ SISSA, Astrophysics Sector, Via Bonomea 265, I-34136 Trieste, Italy \\
$^4$ UPMC Univ Paris 06, Institut d'Astrophysique de Paris, 98 bis,
boulevard Arago 75014 Paris, France \\ $^5$ Department of
Astronomy and Physics, University of Illinois at Urbana-Champaign,
Urbana IL 61801, USA 
}

\begin{document}

\maketitle

\begin{abstract}
  We present a method for the computation of the variance of cosmic
  microwave background (CMB) temperature maps on azimuthally symmetric
  patches using a fast convolution approach. As an example of the application of the method, we show
  results for the search for concentric rings with unusual variance in
  the 7-year \emph{WMAP} data. We re-analyse claims concerning the
  unusual variance profile of rings centred at two locations on the
  sky that have recently drawn special attention in the context of the
  conformal cyclic cosmology scenario proposed by
  \citet{penrose:2009}. We extend this analysis to rings with larger
  radii and centred on other points of the sky. Using the fast
  convolution technique enables us to perform this search with higher
  resolution and a wider range of radii than in previous studies.  We
  show that for one of the two special points rings with radii larger
  than 10 degrees have systematically lower variance in comparison to
  the concordance $\Lambda$CDM model predictions. However, we show
  that this deviation is caused by the multipoles up to order
  $\ell=7$.  Therefore, the deficit of power for concentric rings with
  larger radii is yet another manifestation of the well-known
  anomalous CMB distribution on large angular scales.
  Furthermore, low variance rings can be easily found centred on other
  points in the sky.  In addition, we show also the results of a
  search for extremely high variance rings. As for the low variance
  rings, some anomalies seem to be related to the anomalous distribution of the low-order
  multipoles of the \emph{WMAP} CMB maps. As such our results are not
  consistent with the conformal cyclic cosmology scenario.
\end{abstract}


\section{Introduction}
A wide range of cosmological projects carried out in the last two
decades -- observation of supernova, galaxy surveys and measurements
of cosmic microwave background (CMB) fluctuations -- have resulted in
high quality data enabling the establishment of a standard
cosmological model i.e.~the $\Lambda$CDM model. This model is based on
the Big Bang paradigm formulated in the last century by
\citet{alpher:1948} and extended by the inflationary paradigm by
\citet{guth:1981}. It explains the observed large scale structure of
the Universe as the result of the growth of primordial inhomogeneities
in an expanding Universe, generated during an inflationary epoch, and
driven by the mechanism of gravitational instability. The rate of
expansion and the growth of inhomogeneities depends on the relative
amounts of the different forms of matter and energy that constitute
the total content of the Universe. In the present epoch the Universe
consists of 5\% of baryonic matter, 22\% of dark matter and 73\% of
dark energy \citep{larson:2011}. With only a modest number of free
parameters, the $\Lambda$CDM model is able to successfully fit
thousands of observational data points.

However, this model can not be considered as a fundamental model. It
does not explain the nature of dark matter nor dark
energy. Furthermore, it postulates the existence of an unknown scalar
field responsible for the inflation. Thus, there is a need for
stringent tests of the validity of this model by comparison with
alternative cosmological theories having different observational
predictions.

One example of such an approach is the search in CMB maps for features
that are not predicted by an isotropic and homogeneous standard
cosmological model. Remarkably, this effort has resulted in several
reports of both non-Gaussianity and evidence for the breakdown of
statistical isotropy in the \emph{Wilkinson Microwave Anisotropy
  Probe} (\emph{WMAP}) CMB data, as established by many qualitatively
different methods. In particular, an intriguing planarity and
alignment of the quadrupole and octopole has been demonstrated
\citep{de
  Oliveira-Costa:2004,copi:2004,schwarz:2004,bielewicz:2005,land:2005}.
A localised source of non-Gaussianity, in the form of a very cold spot
on the sky of angular scale 10 degrees, has been detected by
\citet{vielva:2004} and \citet{cruz:2005}.  In addition, an asymmetry
in large-scale power as measured in the two hemispheres of a reference
frame close to that delineated by the ecliptic plane was determined by
\citet{eriksen:2004} and \citet{hansen:2004}
\citep[also][]{hansen:2009,hoftuft:2009}.

An alternative approach is to search the CMB maps for specific
patterns that are signatures of alternative cosmological models.  One
example of such signatures are the pairs of circles with matching
distribution of the CMB anisotropy signal predicted for universes with
multiconnected topology \citep{cornish:1998}. If the size of the
fundamental domain were smaller than diameter of the observable
Universe then we should see at least one pair of such matched
circles. Such searches, as most recently undertaken by
\citet{bielewicz:2011}, have yielded negative results \citep[see
also][]{key:2007}.  However, as pointed out by \citet{bielewicz:2012},
with the upcoming Planck data we should be able to perform a similar
search using CMB polarisation data as a cross-check of the
temperature-based results \citep[see also][]{riazuelo:2006}.

A second example is provided by models of pre-inflationary massive
particles \citep{fialkov:2010}. These predict the existence of
concentric cold and hot rings in the CMB and a bulk flow of galaxies
toward their center. The detection of such rings for the \emph{WMAP}
maps at 3$\sigma$ confidence level was reported recently by
\citet{kovetz:2010}.

Another example of a model that imprints circular patterns on the CMB
sky is due to the theory of eternal inflation \citep{feeney:2011}.
Here, collisions of bubble universes leave signatures in the CMB maps
with a characteristic profile. Hints of evidence of such signatures in
the \emph{WMAP} data were recently reported by \citet{feeney:2011} and
\citet{mcewen:2012}.

Finally, the scenario of conformal cyclic cosmology proposed by
\citet{penrose:2009} \citep[see also][]{penrose:2008,penrose:2010}
proposes that the collisions of massive black holes in a previous
cycle of the Universe results in a set of concentric rings with low
variance on the CMB sky. In this context, rings centred at points with
the Galactic coordinates $(l,b)=(105^\circ,37^\circ)$ and
$(l,b)=(252^\circ,-31^\circ)$ have recently merited special attention
\citep{hajian:2011,moss:2011,wehus:2011}.

In this paper, we perform a complete search for rings with anomalously
variances in the \emph{WMAP} CMB maps. For the computation of
variance, we use an approach based on fast convolution on the sphere
thus enabling the search for concentric rings 
centred on a high resolution grid and with radii in the range from
$0^\circ$ to $90^\circ$. This significantly extends previous studies
that were limited, due to the use of pixel based methods, to rings
centred on a lower resolution grid and radii in the range
$[0^\circ,20^\circ]$. Furthermore, in addition to the low variance
rings we search also for the high variance rings. Though, they are not
predicted by the conformal cyclic cosmology scenario, their presence
would manifestly contradict the standard inflationary models.

It should be noted that \citet{mcewen:2012} used a similar approach in
the context of searching for signatures of eternal inflation.

\section{Data} \label{sec:data}
The search for extreme variance rings was performed on the 7-year
\emph{WMAP} data \citep{jarosik:2011}. The \emph{WMAP} satellite
observes the sky in five frequency bands denoted \emph{K}, \emph{Ka},
\emph{Q}, \emph{V} and \emph{W}, centred on the frequencies of 22.8,
33.0, 40.7, 60.8, 93.5 GHz with angular resolutions of approximately
52.8', 39.6', 30.6', 21' and 13.2', respectively. The
maps\footnote{available at http://lambda.gsfc.nasa.gov} are pixelised
in the \textsc{healpix}\footnote{http://healpix.jpl.nasa.gov} scheme
\citep*{gorski:2005} with a resolution parameter $N_{\rm side}=512$,
corresponding to 3,145,728 pixels with a pixel size of $\sim 7$
arcmin. In particular, we studied the V and W-band maps corrected for
Galactic emission using a template fitting approach \citep{gold:2011}.

\section{Computation of variance by fast convolution on the sphere}
The variance of the CMB in concentric rings with radius $r$
and width $\Delta r$ centred on pixel $i$, ${\rm Var}_{i,r}(\Delta
T)$, can be estimated via the expression: 
\begin{eqnarray} \label{eqn:variance}
{\rm Var}_{i,r}(\Delta T) &=&  \frac{\sum_j \Delta T^2_j M_j
  B^r_{ij}}{\sum_k M_k^{} B^r_{ik}} - \left( \frac{\sum_j \Delta T_j M_j
  B^r_{ij}}{\sum_k M_k^{} B^r_{ik}} \right)^2 \nonumber \\ 
 & & - \frac{\sum_j \sigma^2_j M_j
  B^r_{ij}}{\sum_k M_k^{} B^r_{ik}}\ ,
\end{eqnarray}
where $M_i$ denotes mask applied to the map and $B^r_{ij}$ is the ring profile such that
\begin{equation} \label{eqn:profile}
B^r_{ij} = \left\{ \begin{array}{ll} 
1 & {\rm for}\ r \leq \arccos(\boldsymbol{\hat{n}}_i \cdot \boldsymbol{\hat{n}}_j) < r+\Delta r \\
0 & {\rm otherwise}
\end{array} \right. \ .
\end{equation}
The sum in the denominators is over the unmasked pixels in a given
ring. The last term in equation (\ref{eqn:variance}) corresponds to a noise
variance term that has to be subtracted to yield an unbiased estimation of the
CMB variance. The noise variance in a given pixel $\sigma_i^2$ is defined as
$\sigma_i^2 \equiv \sigma^2_{\rm obs} / N_i^{\rm obs}$, where
$\sigma_{\rm obs}$ is the rms of noise per observation and $N_i^{\rm
  obs}$ denotes number of observation of a given pixel. Both of these
quantities are provided by the \emph{WMAP} team for each map.

The correction for the noise bias is not needed if we estimate the cross-variance
of two differencing assemblies (DAs) for a
given band. Such an estimator is not susceptible to uncertainties in
the modeling of the noise properties. In this case, the estimator of the
variance takes the form:
\begin{eqnarray} \label{eqn:crossvariance}
{\rm Var}_{i,r}(\Delta T) &=&  \frac{\sum_j \Delta T_{1,j} \Delta T_{2,j} M_j
  B^r_{ij}}{\sum_k M_k^{} B^r_{ik}}  + \\
 & & - \frac{\left( \sum_j \Delta T_{1,j} M_j
  B^r_{ij} \right) \left( \sum_j \Delta T_{2,j} M_j
  B^r_{ij} \right) }{\left(\sum_k M_k^{} B^r_{ik} \right)^2} \ , \nonumber 
\end{eqnarray}
where $\Delta T_1$ and $\Delta T_2$ denote maps for two DAs. 
We will use this estimator to derive most of the results
presented below. We checked that the differences between the two
estimators are negligible for the 7-year \emph{WMAP} data. In the
case of the W-band, we used coadded maps for two 
pairs of DAs i.e.~coadded (W1+W2)/2 and (W3+W4)/2 maps.

The computation of the estimator of CMB variance by direct summation
over the pixels within the ring for all ring centers is very time
consuming. However, one can observe that the variance estimator given
by equation (\ref{eqn:variance}) is simply the convolution of the square of the masked map, corrected
for the mean, with the ring profile of a given radius and width.  For
the cross-variance, the product of the two maps is convolved with the
ring profile. Thus, the computations can be significantly speeded up
by use of a technique using fast convolution on the sphere and
therefore performing all essential computations in the spherical
harmonics space. Because the ring profile is
azimuthally symmetric the computations are the same as for convolution
with an azimuthally symmetric beam. In order to estimate the variance,
we need to perform four such convolutions. We used the fast spherical
harmonics transform implemented in the \textsc{healpix} scheme for
these computations.

Such an approach allows us to perform a search for rings with an
unusual variance centred on a higher resolution grid and for a wider
range of ring radii than by using the pixel based methods employed in
previous studies.  It can be also easily extended to higher moments of
the map such as skewness and kurtosis, as well as employed for more
general azimuthally symmetric profiles such as discs with an arbitrary radial
weighting function. In fact,  \citet{bernui:2009,bernui:2010,bernui:2011,bernui:2012}
 have previously computed the skewness and kurtosis of CMB data. 
 However, our approach allows the analysis to be rapidly performed on spherical caps over
  a range of radii and for all positions on the sky.

\begin{figure*}

\centerline{
\epsfig{file=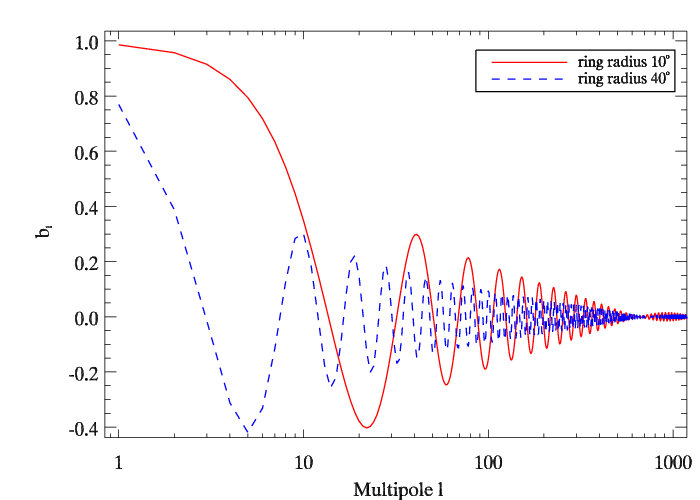,scale=0.4}
}

\caption{The spherical harmonic coefficients $b_\ell$ of rings
  with outer radii of $10^\circ$ (red solid line) and $40^\circ$
  (dashed blue line) and a width of $0.5^\circ$.} 
\label{fig:ring_profile}
\end{figure*}

Examples of the spherical
harmonic window functions for example ring geometries, $b_\ell \equiv
B_{\ell 0} / \sqrt{2 \ell +1}$, are shown in
\fig\ref{fig:ring_profile}. Due to the azimuthal symmetry of the ring,
the spherical harmonics coefficients, 
\begin{equation} \label{eqn:blm}
B_{\ell m} \propto \delta_{m 0} \int_{\cos(r+\Delta r)}^{\cos(r)} P_\ell (\cos \theta)\, {\rm d}\cos \theta \ ,
\end{equation}
where $P_\ell (\cos \theta)$ are the Legendre polynomials, are
non-zero only for modes $m=0$. The factor $\sqrt{2\ell+1}$ comes from
the averaging of the squares of the coefficients over $m$. The window
functions were normalised to one at $b_0$. Note that, due to
pixelisation effects, the ring window function corresponding to a
pixelised sphere is not exactly equal to the analytical expression for
the window function given by relation (\ref{eqn:blm}).
  Moreover,  the actual ring profile should vary over the sky 
  due to both the orientation of the ring with respect to the coordinate
  frame and variation in shape of the \textsc{healpix} pixels.
  However, this effect is small and its impact on the analysis is negligible.
  In these computations, we utilise the window functions determined for
  the set of pixelised rings of given radii and widths centred on the pole.

As can be seen, the ring window function oscillates around zero with a
wavelength $\sim 2\pi/r$ dependent on the ring radius $r$ expressed in
radians i.e.~ 36 and 9 for rings of width $0.5^\circ$ with outer radii
of $10^\circ$ or $40^\circ$, respectively. Furthermore, because of the
$\sqrt{2 \ell +1}$ factor in the denominator, the envelope of the
widow function falls as $\ell^{-1/2}$. Additional modulation of the
envelope depends on the width of the ring $\Delta r$. To better
understand this behaviour, the ring profile can be considered as a
disc of given outer radius from which a smaller disc of the given
inner radius is subtracted. The interference of the window functions
with slightly different wavelengths corresponding to these discs
introduces a modulation with the beat wavelength $\sim 2\pi / \Delta
r$. For the ring width of $0.5^\circ$ the wavelength is around 720
which corresponds to the node of the window functions seen in the
figure. Note also that a larger fraction of the total power for rings
with a larger radius comes from the lowest multipoles in comparison to
rings with a smaller radius. Thus, the former are more sensitive to
the large angular scale power of the CMB maps. This dependence will be
clearly seen in the different tests presented in the paper.

Finally, it is worth pointing out that one can consider a more general
estimator than that given by equations (\ref{eqn:variance}) or
(\ref{eqn:crossvariance}). The terms in the sum can be weighted with
coefficients $w_i$ constrained by additional conditions imposed on the
estimator. In the case of an optimal estimator, in the sense of
minimisation of its variance, the weights are given by $w_i =
1/(S+\sigma_i^2)^2$ where \mbox{$S \equiv 1/ (4 \pi) \sum_\ell (2 \ell
  +1)\, C_\ell \, B_\ell^2$} denotes the average CMB variance for the
sky observed with an azimuthally symmetric beam characterised by the
window function $B_\ell$.  Nevertheless, for a high signal-to-noise
ratio, the variance of the optimal estimator is almost the same as for
the suboptimal one. This is the case for the 7-year \emph{WMAP}
data. While the CMB variance for the V-band map is about $7800\ \mu
{\rm K}^2$, the median noise variance in a pixel is about $3300\ \mu
{\rm K}^2$. Indeed, a comparison between the optimal and suboptimal estimator for
the \emph{WMAP} data and found only negligible
differences. Furthermore, the optimal estimator requires additional
computing time since two additional convolutions are performed.  For
these reasons, we decided to used the suboptimal estimator in our
analysis.

\section{Statistics} \label{sec:statistic}
Here we specify the two statistics used to test the significance of
the observed variance profiles -- the $\chi^2$ statistic and the number of rings
with extreme variance values.

\subsection{$\chi^2$ statistic} \label{sec:chisq_test}

In order to test the significance of the measured variances for
concentric rings centred on a given pixel $i$, we used the $\chi^2$
statistic defined as
\begin{equation} \label{eqn:chisq}
\chi^2_i = \sum_{r,r'} \left( {\rm Var}_{i,r} - \mu_{i,r} \right)
C^{-1}_{i,rr'} \left( {\rm Var}_{i,r'} - \mu_{i,r'} \right)\ ,
\end{equation}
where $\mu_{i,r}$ and $C_{i,rr'}$ are the mean variance and covariance
matrix of the variances for a given pixel $i$, respectively. Note that
the application of a mask results in a dependence of the mean and
covariance matrix values on the position on the sky of the ring.
Consequently, the computation of the $\chi_i^2$ statistic becomes
complicated. To store all of the covariance matrices $C_{i,rr'}$
requires of order $N_{\rm pix} N_{\rm bins}^2$ bytes of disk
space. For maps with $N_{\rm pix}$ pixels on the sky corresponding to
a resolution parameter $N_{\rm side} = 512$, and to a number of bins
$N_{\rm bins} = 180$, corresponding to rings of width $0.5^\circ$ with
a radius in the range 0 to 90 degrees, this corresponds to
approximately 400 GB of disk space. Thus to make these computations
feasible, after convolution of the original map
with the ring window function in spherical harmonic space we 
construct a variance map at $N_{\rm side}=128$. This is sufficient
resolution to probe the variance distribution in the sky for rings of
width $0.5^\circ$. Furthermore, we divided the sum over bins of ring
radii into three intervals: $[0^\circ,30^\circ[$,
$[30^\circ,60^\circ[$ and $[60^\circ,90^\circ[$. The required disk
space is then reduced to around 2.8 GB.

The conformal cyclic
  cosmology scenario does not provide well established predictions for
  the width of the rings that are being searched for. However,
  assuming that the typical size is proportional to
  the ratio of the collision time of the massive black holes to the
  total time of the previous cycle of the Universe, one can predict
  that the rings should be very narrow with a width rather below the 
  angular resolution of the \emph{WMAP} maps. Therefore, in this
  analysis, we only consider rings of width $0.5^\circ$,  only
  slightly larger than the angular resolution ($\sim 20'$)  of the V-band data
  used predominantly in our analysis. In addition,  this choice allows for
  a straightforward comparison of our results with previous studies.
  Note that there are also computational
  constraints on the width of the rings since an analysis with a narrower
  width is more computationally demanding. 

The mean and covariance matrices were estimated on the basis of 1000
Monte Carlo (MC) simulations of the 7-year \emph{WMAP} maps for the
standard $\Lambda$ CDM cosmological model \citep[see][Table
3]{larson:2011}. Another set of simulations allowed the estimation of
the significance of the observed values of the $\chi_i^2$ statistic.
The $\chi_i^2$ statistic was not computed for any ring for which more
than 50\% of the component pixels were masked by the \emph{WMAP}
KQ75y7 mask. Such rings are delineated by masking the corresponding pixel defining
the centre of the ring. Masks for the $\chi_i^2$ maps determined in
this way are shown in the figures discussed below.

\subsection{Number of extreme variance rings} \label{sec:count_extremea}

Another statistic which we used was the number of rings centered on a
given pixel with extremely low or high variance. Extreme values were
defined to be those deviating from the mean by more than one and a
half standard deviations. To estimate the significance of the observed number of extreme rings
we use MC simulations of the CMB maps and p-values for the null
hypothesis i.e.~the probability of getting a larger number of extreme
rings for the standard $\Lambda$CDM cosmological model. As was the
case for the $\chi^2$ statistic, we divided the ring radii into three
intervals: $[0^\circ,30^\circ[$, $[30^\circ,60^\circ[$ and
$[60^\circ,90^\circ[$. If concentric rings of different radii centred
at a given point are independent, then the p-value for the number of
extreme variance rings $n$ can be estimated from the cumulative
binomial distribution
\begin{equation}
P(n) =\sum_{k > n}^{N_{\rm bins}} {N_{\rm bins} \choose k} p^k\, (1-p)^{N_{\rm bins}-k} \ ,
\end{equation}
where $p$ is the probability of getting an extreme variance for a
given ring and $N_{\rm bins} = 60$ for one of the three radius intervals of rings
with a width of $0.5^\circ$. We checked that this approximation of the
p-value is sufficiently valid for the radius interval
$[0^\circ,30^\circ[$. However, it is not valid for the other intervals
because of strong correlations between rings of larger radii.

\section{Results}
\begin{figure*}

\centerline{
\epsfig{file=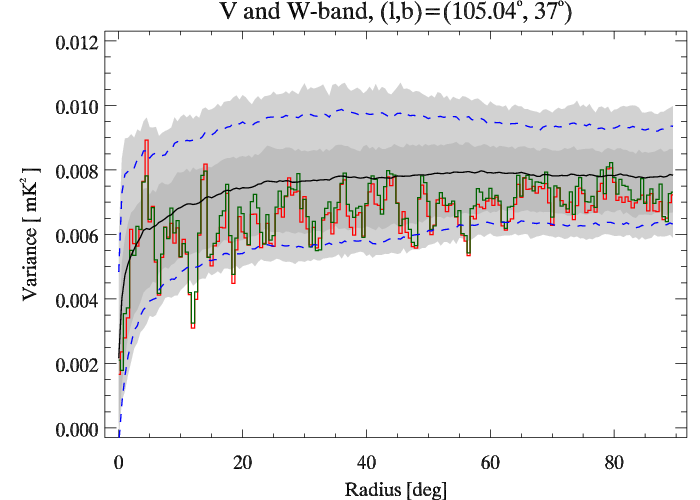,scale=0.26}
\epsfig{file=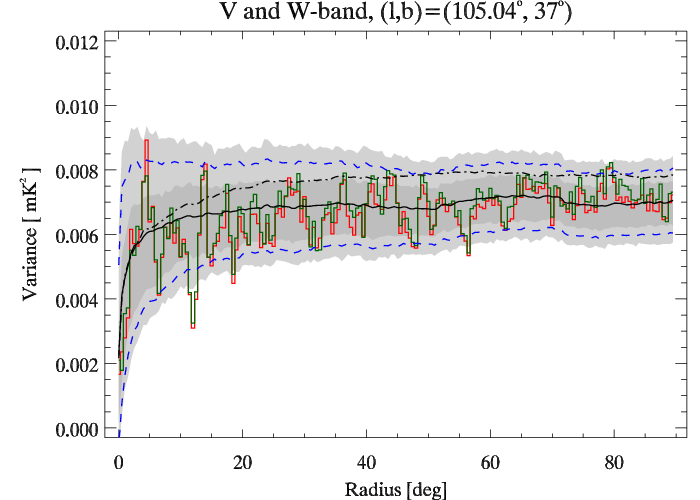,scale=0.26}
\epsfig{file=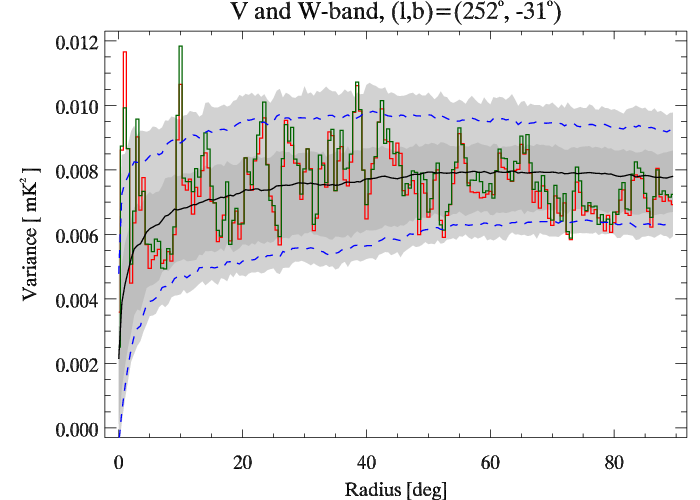,scale=0.26}
}

\caption{Variance for rings centred at the Galactic coordinates
  $(l,b)=(105^\circ,37^\circ)$ (left and middle) and
  $(l,b)=(252^\circ,-31^\circ)$ (right) with radii in the range
  $[0^\circ,90^\circ[$. The red and green lines denote the variance for 
the V and W-band maps, respectively, masked with the KQ75y7 cut. The
black lines denote the mean variance estimated from 1000
simulations. The dashed blue lines correspond to the mean shifted up and
down by one and a half standard deviations of the variance. The shaded
dark and light grey regions
indicate the 68\% and 95\% confidence regions, respectively. In the middle figure, the mean
and confidence regions were estimated using simulations with 
multipoles in the range $[0,7]$ replaced by the corresponding best-fitting
multipoles from the foreground corrected V-band \emph{WMAP} map. The
dash-dotted line in this figure denotes the mean for simulations without replacing the
lowest order multipoles i.e.~the same mean as in the left figure.
Note that the confidence regions are narrower than 
for the simulations which do not replace the lowest order
multipoles. This should not be surprising because the cosmic variance for
simulations with fixed low-order multipoles must be smaller.
}
\label{fig:variance_profile}
\end{figure*}

Before presenting results of the analysis for the full sky, we would like to
first address the properties of the sets of concentric rings centred on two
points which have recently drawn special attention from the CMB
community. This interest was based on claims related to the
statistical significance of the variances measured on these rings in
the context of the conformal cyclic cosmology scenario proposed by
\citet{penrose:2009}.
The variance profiles for the rings
centred at the Galactic coordinates $(l,b)=(105^\circ,37^\circ)$
and $(l,b)=(252^\circ,-31^\circ)$ for ring
radii from zero to 90 degrees are shown in
\fig\ref{fig:variance_profile}. Comparing the observed profiles with
simulations based on the standard $\Lambda$CDM model we do not identify any significant number
of unusually extreme values of the variance in good agreement
with the conclusions in \citet{hajian:2011,moss:2011,wehus:2011}.
However, since our analysis is not
limited only to a range of radii from 0 to 20 degrees as in the previous works,
we can extend this conclusion to the full range of ring radii from 0 to 90 degrees.

Nevertheless, for the point $(l,b)=(105^\circ,37^\circ)$ there is
evidence of systematically smaller variances for concentric rings with
radii greater than $10^\circ$. Since such rings are more sensitive to
larger angular scales, we consider that the lower variance might be
related to the large angular scale anomalies observed in the
\emph{WMAP} data, and in particular to the hemispherical asymmetry in
the power spectrum \citep{eriksen:2004}. Specifically, the point
$(l,b)=(105^\circ,37^\circ)$ lies in the northern ecliptic hemisphere
which exhibits less power than the southern hemisphere. To test this
hypothesis, we performed additional simulations in which the lowest
multipoles ($\ell \leq 7$) were replaced by the corresponding best-fitting
values determined from the foreground corrected V-band data for the
KQ75y7 sky coverage. 

Specifically, the low-order multipoles were reconstructed from
  cut sky data using the direct inversion method described in detail in
  the appendix \ref{sec:appendix}. Although this approach is less optimal than 
  the maximum likelihood \citep{deoliveira:2006} or Wiener filtering
  \citep{tegmark:1997} methods, it provides a
  reliable and unbiased estimation of the low-order 
  multipoles \citep{bielewicz:2004,efstathiou:2004}.
  Hereafter, we will refer to the
  low-order multipoles reconstructed in this way also as the best-fitting multipoles.

 \fig\ref{fig:variance_profile} shows that
simulations with replaced best-fitting multipoles are more consistent with the data. Replacing fewer
multipoles did not reconcile the anomalous behaviour fully, and
certainly substituting only the multipoles up to order $\ell=3$ is not
sufficient, suggesting that the large scale anomalies are not limited
only to the apparently aligned quadrupole and octopole.

It is important to note that that a systematically smaller variance
for rings with larger radii is not consistent with the conformal
cyclic cosmology scenario. In this model, low variance rings are
generated by the collisions of massive black holes in the previous
cycle of the Universe. For each of the collisions, one should observe
a low variance ring centred at the position of a given black hole. As
the black hole undergoes a series of collisions during its lifetime,
we should observe a series of concentric low variance rings with
different radii. Earlier collisions will generate rings with a larger
radius while later ones result in rings with a smaller
radius. However, for a significant fraction of its lifetime, a black
hole will not modify the statistical properties of the CMB sky and
rings not related to the collisions will show typical values of the
variance. The fact that we find only a few rings with variance bigger
than the mean for radii larger than $10^\circ$ is not consistent with
these predictions. As we showed above, the systematically lower
variance is much better explained by the anomalous distribution of the
low-order multipoles.

\begin{figure*}

\centerline{
\epsfig{file=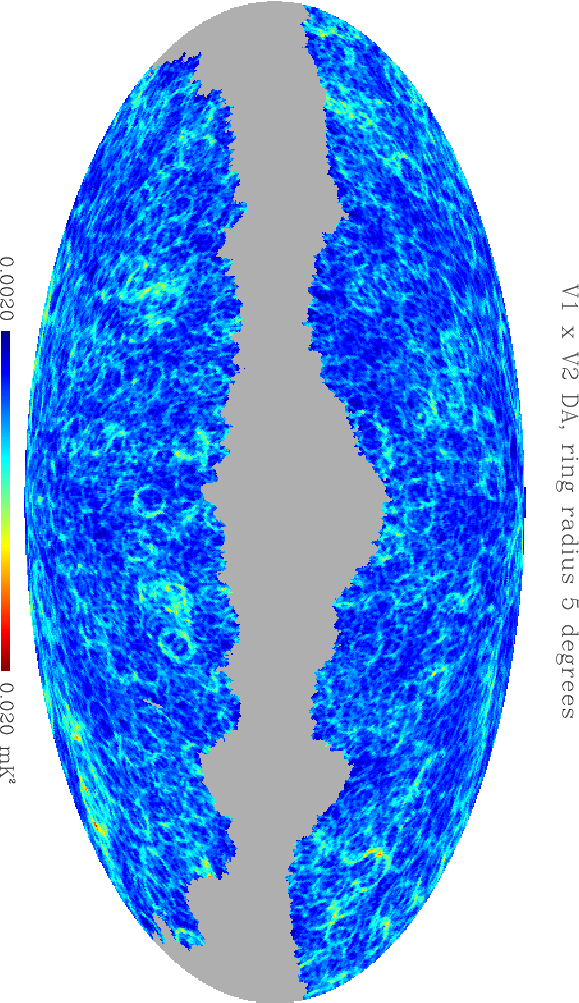,scale=0.23,angle=90}
\epsfig{file=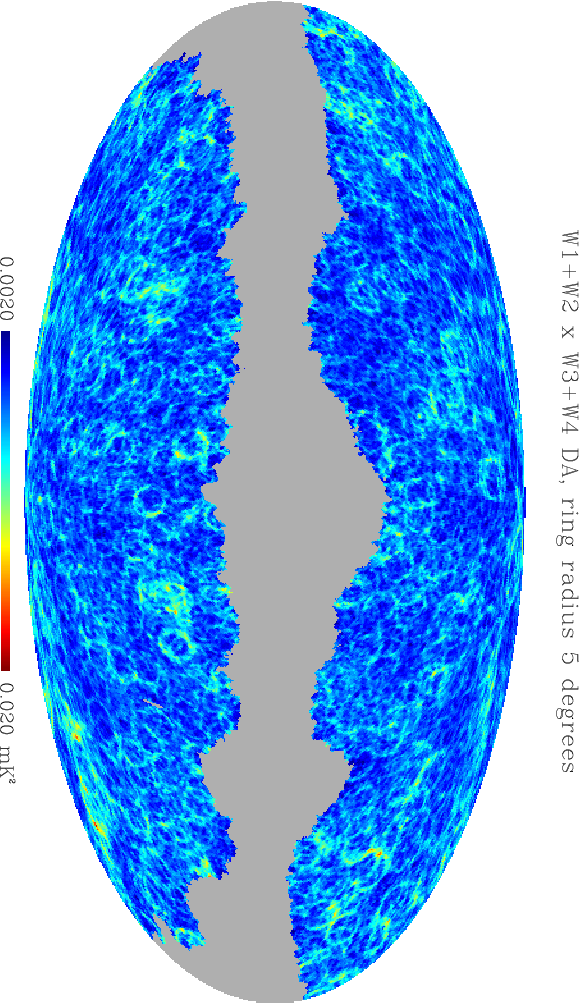,scale=0.23,angle=90}
}

\caption{An example of a variance map for rings with a radius of
  $5^\circ$.  The left figure shows the variance for the
  cross-correlation of the V1 and V2 DAs computed after application of
  the KQ75y7 mask. The right figure shows the analogous map determined
  from the cross-correlation of two pairs of W-band DAs -- (W1+W2)/2
  and (W3+W4)/2.  The grey region corresponds to the pixels for which
  the variance was not computed due to the mask.}
\label{fig:variance_map}
\end{figure*}

In \fig\ref{fig:variance_map} we show examples of the variance maps
for a ring radius of 5 degrees. In appearance, the maps resemble a
collection of overlapping rings of different amplitude. The high
amplitude rings are due to pixels with lower or higher values in the
original map which also form part of the rings with centres separated
by $5^\circ$. It is clear that the variance maps are very similar for
both V- and W-band maps. Since we also did not find significant
differences between the results for the two bands in the statistics
computed below, in the reminder of the paper we show results only for
the V-band map.

\subsection{$\chi^2$ statistic} \label{sec:chisq_test_result}

\begin{figure*}

\centerline{
\epsfig{file=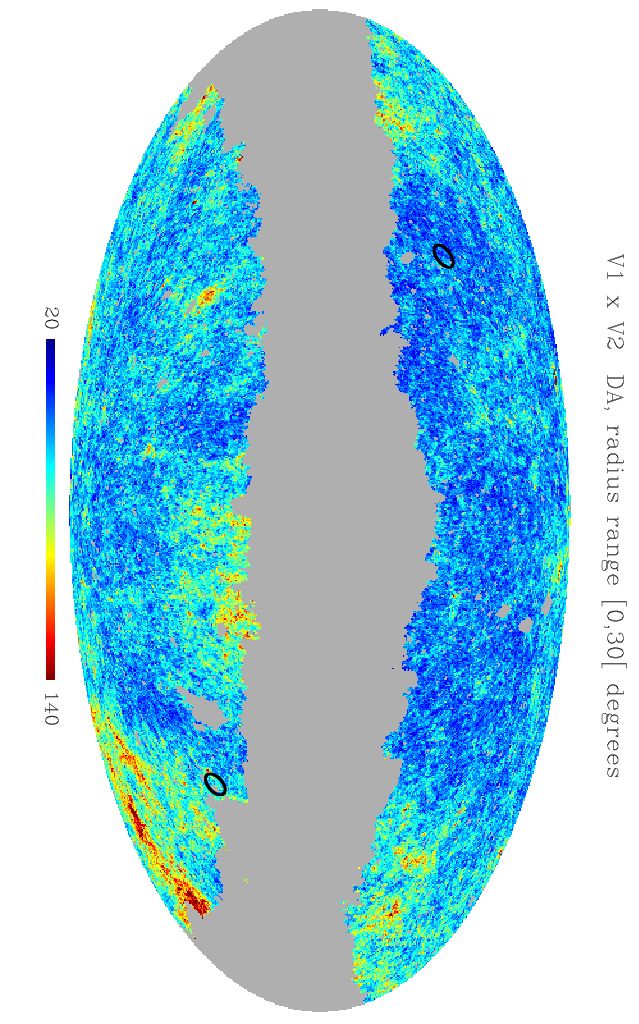,scale=0.18,angle=90}
\epsfig{file=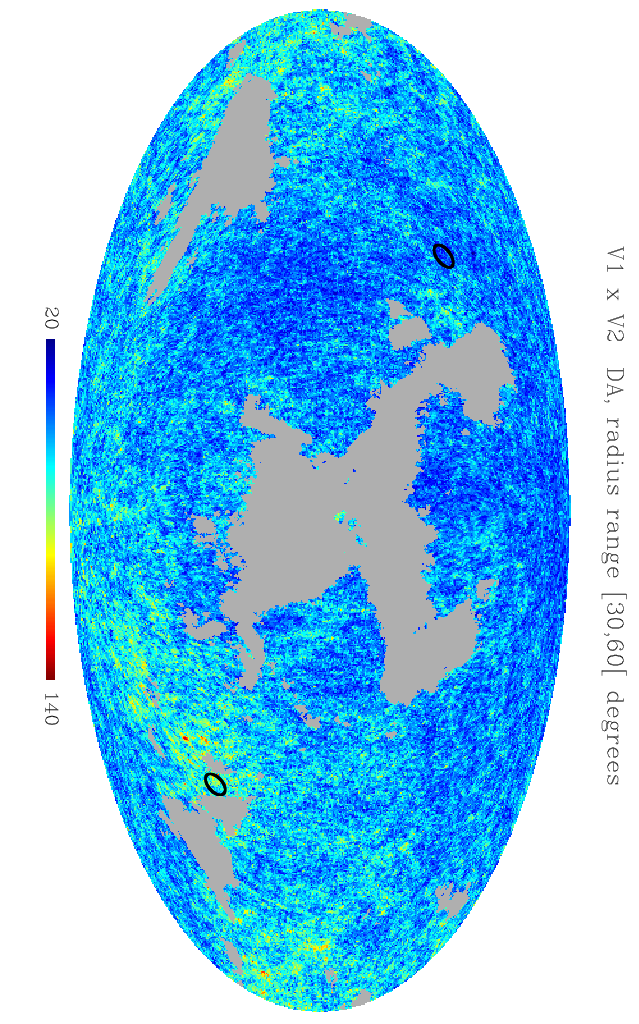,scale=0.18,angle=90}
\epsfig{file=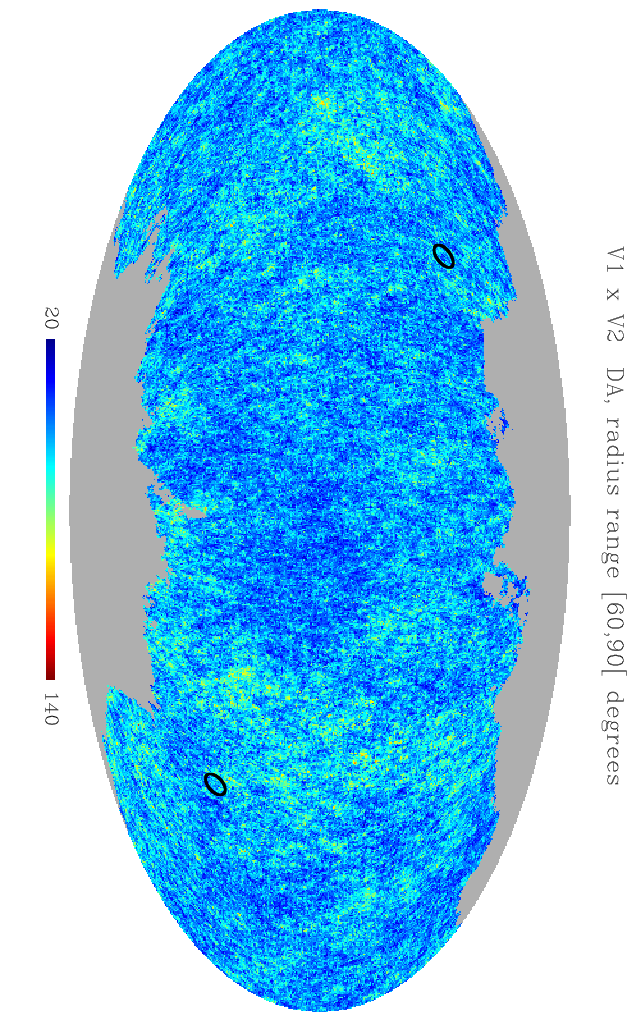,scale=0.18,angle=90}
}

\caption{The $\chi^2_i$ map for the three
  intervals of the ring radii: $[0^\circ,30^\circ[$ (left column) ,
  $[30^\circ,60^\circ[$ (middle column) and
  $[60^\circ,90^\circ[$ (right column). The maps correspond to the
  cross-correlation of the V1 and V2 DAs
  masked  by the KQ75y7 mask. The black circles are centred at two
  points analysed in Fig.~\ref{fig:variance_profile}. The Grey region
 corresponds to pixels for which the $\chi_i^2$ was not computed due
 to the applied mask.} 
\label{fig:chisq_map}
\end{figure*}

The $\chi^2$-maps for three radii intervals are shown in
\fig\ref{fig:chisq_map}. The grey regions correspond to pixels for
which the $\chi^2$ statistic was not computed since at least one of
the concentric rings centered on a given pixel was masked by more than
50\%. It is worth noting that in the case of the intervals
$[30^\circ,60^\circ[$ and $[60^\circ,90^\circ[$ the $\chi^2$ statistic
is computed for those rings centred on some pixels that are located
inside the KQ75y7 mask.

There are several regions on the sky with high values of $\chi^2$ for
the ring radii in interval $[0^\circ,30^\circ[$. However, it is rather
unusual that there is a lack of such regions for the $\chi^2$ maps
corresponding to rings with radii in the intervals $[30^\circ,60^\circ[$
and $[60^\circ,90^\circ[$.

\begin{figure*}

\centerline{
\epsfig{file=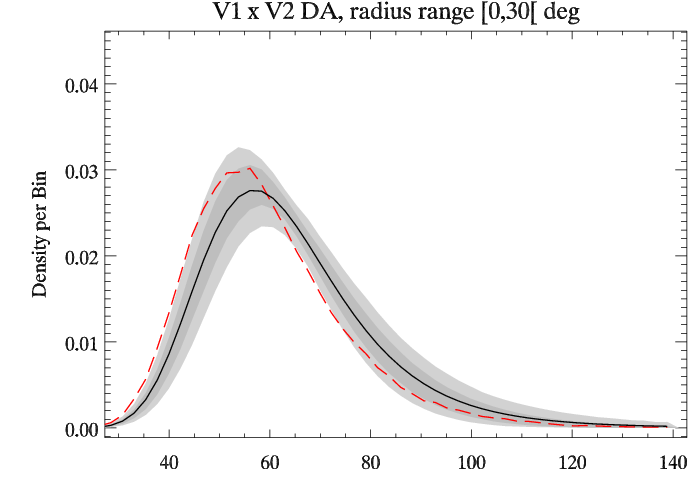,scale=0.25}
\epsfig{file=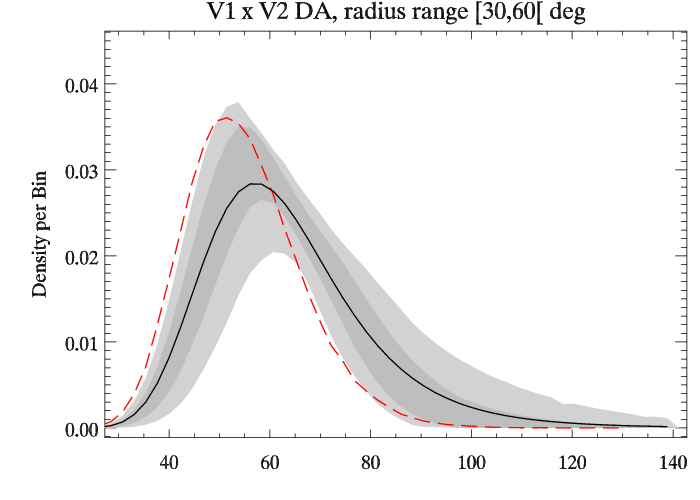,scale=0.25}
\epsfig{file=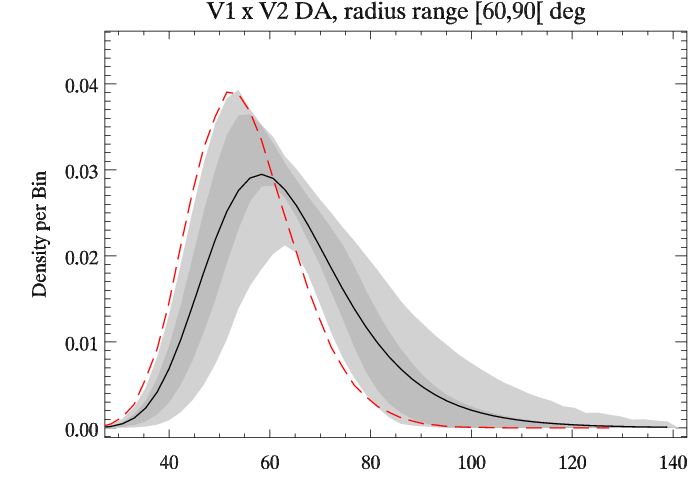,scale=0.25}
}

\caption{The distribution of the $\chi^2_i$ map for the three
  intervals of the ring radii: $[0^\circ,30^\circ[$ (left) , $[30^\circ,60^\circ[$ (middle) and
  $[60^\circ,90^\circ[$ (right). The red lines denote the distribution for
the \emph{WMAP} maps masked with the KQ75y7 mask and the black lines
indicate the mean
of the distribution derived from 1000 MC simulations of signal plus
noise. The shaded dark and light grey regions
indicate the 68\% and 95\% confidence regions, respectively. The distributions correspond to the
cross-correlation of the V1 and V2 DA maps. 
}
\label{fig:chisq_dist}
\end{figure*}

\begin{figure*}

\centerline{
\epsfig{file=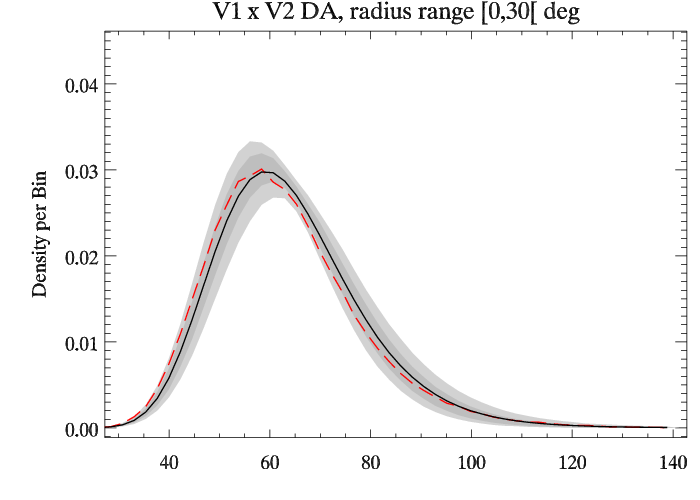,scale=0.25}
\epsfig{file=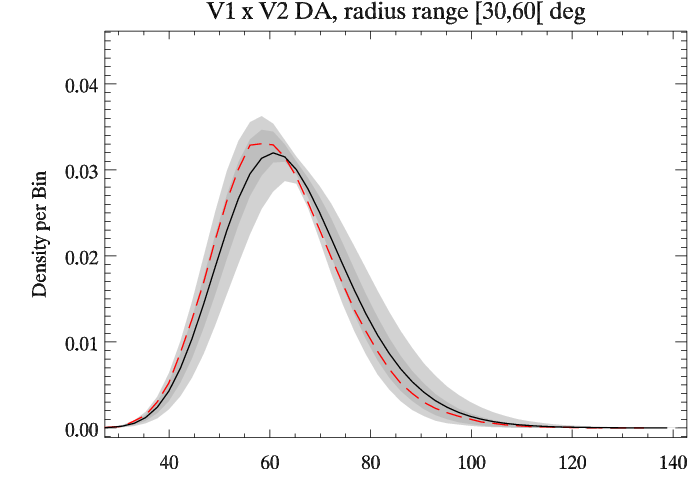,scale=0.25}
\epsfig{file=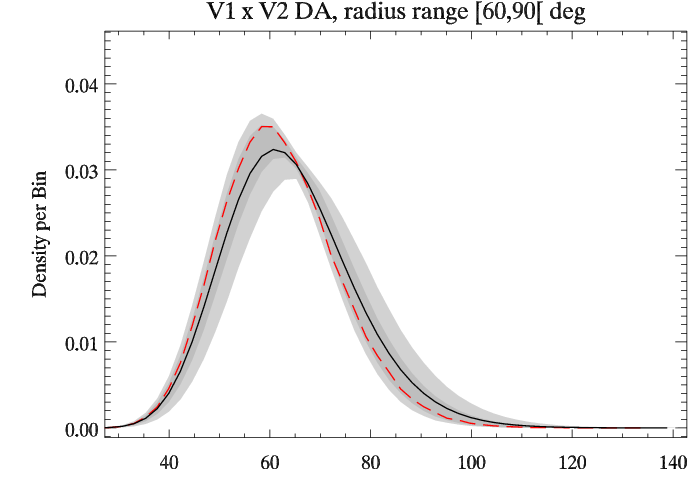,scale=0.25}
}

\caption{The distribution of the $\chi^2_i$ map after removal of the
  best-fitting multipoles in the range $\ell \in  [0,7]$ from the V1 and
  V2 DAs. Plotted lines as in Fig.~\ref{fig:chisq_dist} 
}
\label{fig:chisq_dist_l7removed}
\end{figure*}

To test the significance of the features observed in the $\chi^2$ maps,
we used 1000 MC simulations of the \emph{WMAP} maps. The distributions
of the $\chi^2$ maps are shown in \fig\ref{fig:chisq_dist}. 
If the variances for the different rings were uncorrelated, then
the number of degrees of freedom would be equal to the number of bins
used in computation of the $\chi^2$ i.e.~60 for each of the analysed
ring radius intervals. However, due to correlations, the effective
number of degrees of freedom will be smaller than 60, and this is
apparent from the maximum of the observed distribution.
The distribution for the data lies on the border of the 95\%
confidence region for rings with radii in the intervals
$[30^\circ,60^\circ[$ and $[60^\circ,90^\circ[$.  For the interval
$[0^\circ,30^\circ[$, the data is more consistent with the assumed
$\Lambda$CDM model. Since rings with larger radii are more sensitive
to large angular scales, it is plausible that the observed deviation
between data and theory is caused by large angular scale anomalies. To
test this hypothesis, we repeated the V-band analysis for the data
from which the the best-fitting multipoles in the range $\ell \in [0,7]$
were removed. \fig\ref{fig:chisq_dist_l7removed} shows that this
significantly improves the consistency of the data with the model.

To quantify the difference between the $\chi^2$ distributions
  recovered from the \emph{WMAP} data and the averaged distribution
  determined from the MC simulations we computed the Kolmogorov-Smirnov statistic,
  with results presented in Table \ref{tab:ks_stat}. The distance between
  distributions is significantly smaller after removal of the low-order multipoles.

\begin{table}
\begin{center}
\caption{The Kolmogorov-Smirnov statistic for the cumulative distributions of
  the $\chi_i^2$ maps presented in Fig.~\ref{fig:chisq_dist} (upper
  row) and \ref{fig:chisq_dist_l7removed} (bottom row).} 
\label{tab:ks_stat}
\begin{tabular}{|c|ccc|}\hline
 Radius range            &   $[0^\circ,30^\circ[$ & $[30^\circ,60^\circ[$ &  $[60^\circ,90^\circ[$ \\ \hline
low-$\ell$ included  & 0.110 & 0.231 & 0.249 \\
low-$\ell$ removed  & 0.040 & 0.057 & 0.065 \\
\hline				  		      	    		  
\end{tabular}			  		      	    	  
\end{center}
\end{table}

\begin{figure*}

\centerline{
\epsfig{file=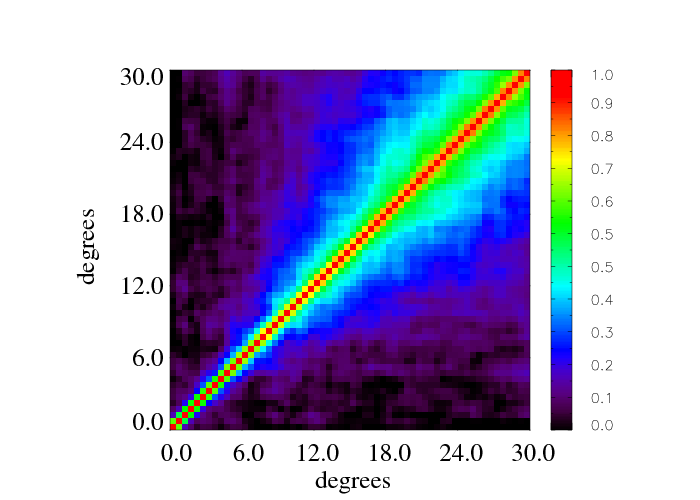,scale=0.26}
\epsfig{file=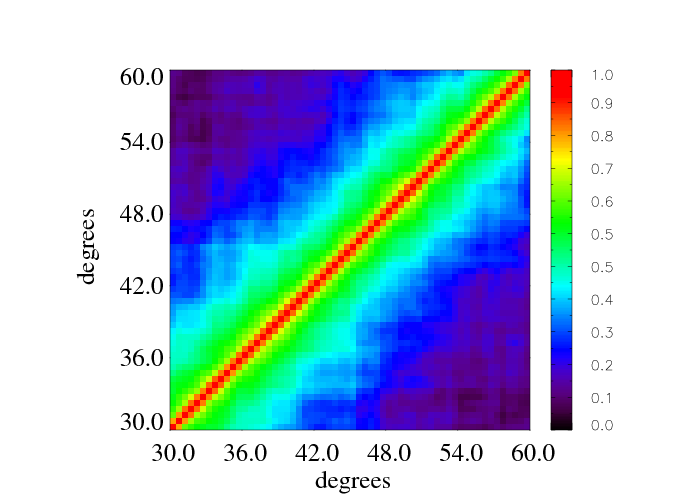,scale=0.26}
\epsfig{file=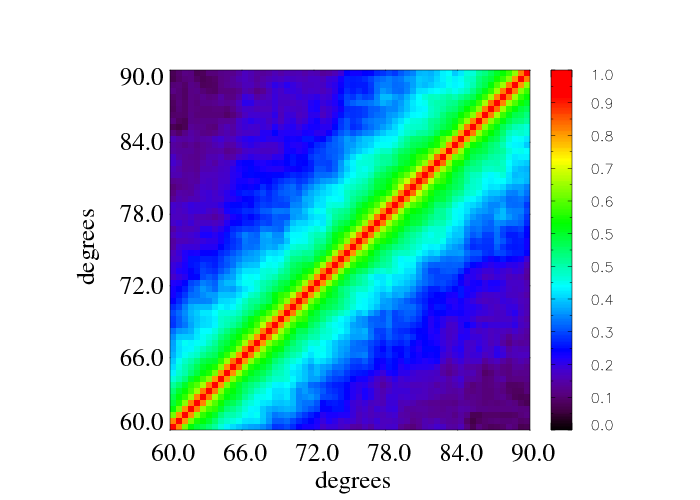,scale=0.26}
}

\centerline{
\epsfig{file=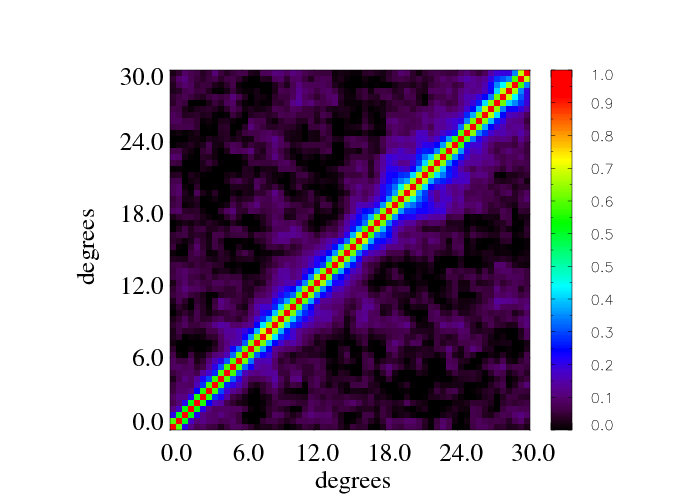,scale=0.26}
\epsfig{file=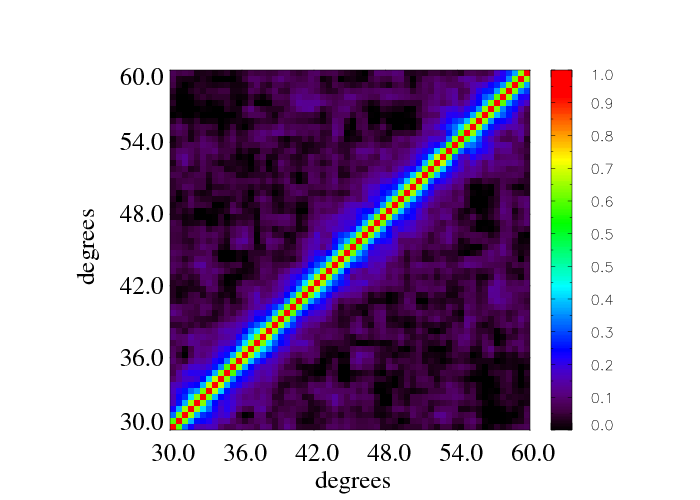,scale=0.26}
\epsfig{file=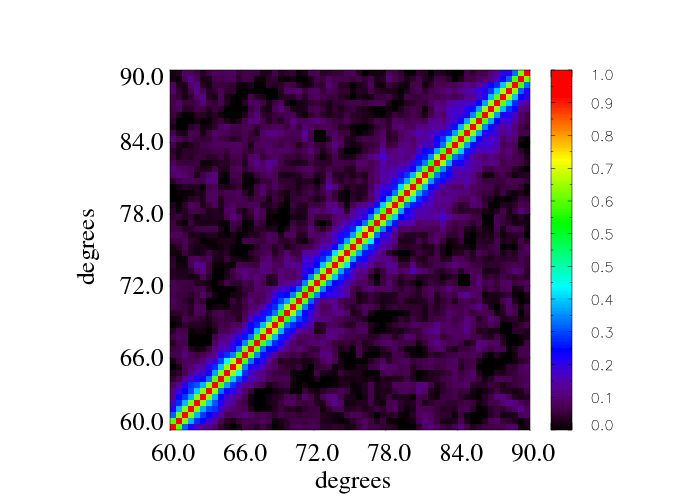,scale=0.26}
}

\caption{Correlation matrices for the variance on concentric rings centred
at the Galactic coordinates $(l,b)=(105^\circ,37^\circ)$ and ring
radius intervals $[0^\circ,30^\circ[$ (left column) ,
$[30^\circ,60^\circ[$ (middle column) and
  $[60^\circ,90^\circ[$ (right column).
The bottom row corresponds to the matrices derived after removing the best-fitting multipoles in
the range $\ell \in  [0,7]$ from the maps. The matrices were estimated using 1000 MC simulations.} 
\label{fig:corr_mat}
\end{figure*}

Correlations in the variance between rings of different radius are
shown in \fig\ref{fig:corr_mat}. The correlation matrices correspond to rings centred on the Galactic
coordinates $(l,b)=(105^\circ,37^\circ)$ for the three ring radius
intervals and are normalised by the diagonal terms of the covariance
matrix used in the computation of the $\chi_i^2$
(Eqn.~\ref{eqn:chisq}). Rings with a larger radius are strongly
correlated. This is seen particularly well for the intervals
$[30^\circ,60^\circ[$ and $[60^\circ,90^\circ[$. However, removing
from the map the best-fitting multipoles in the range $\ell \in [0,7]$
significantly decorrelates the rings.

\subsection{Number of extreme variance rings} \label{sec:count_extremes_result}

\begin{figure*}

\centerline{
\epsfig{file=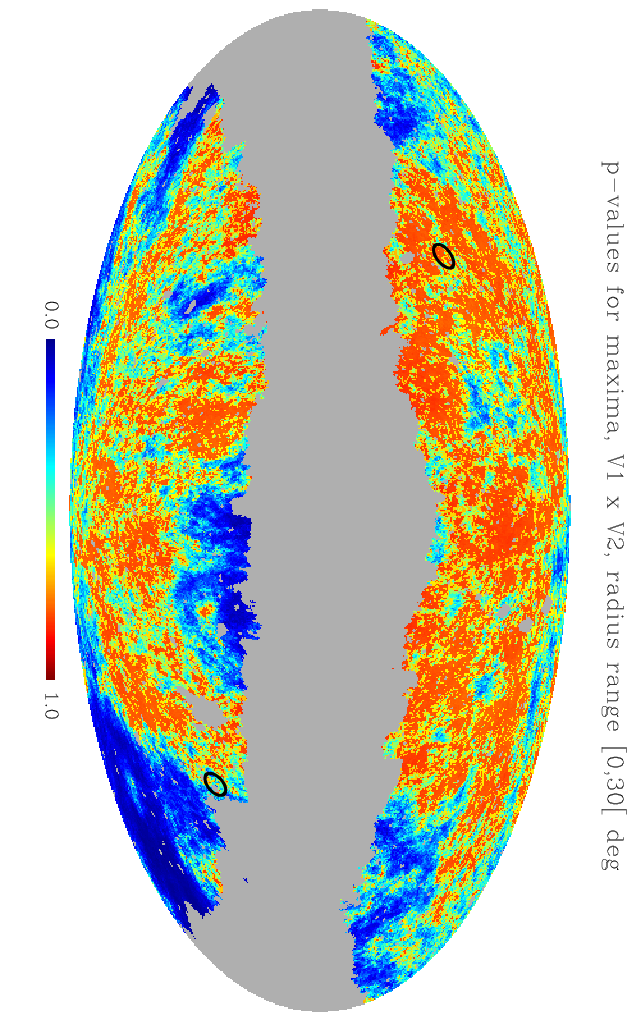,scale=0.18,angle=90}
\epsfig{file=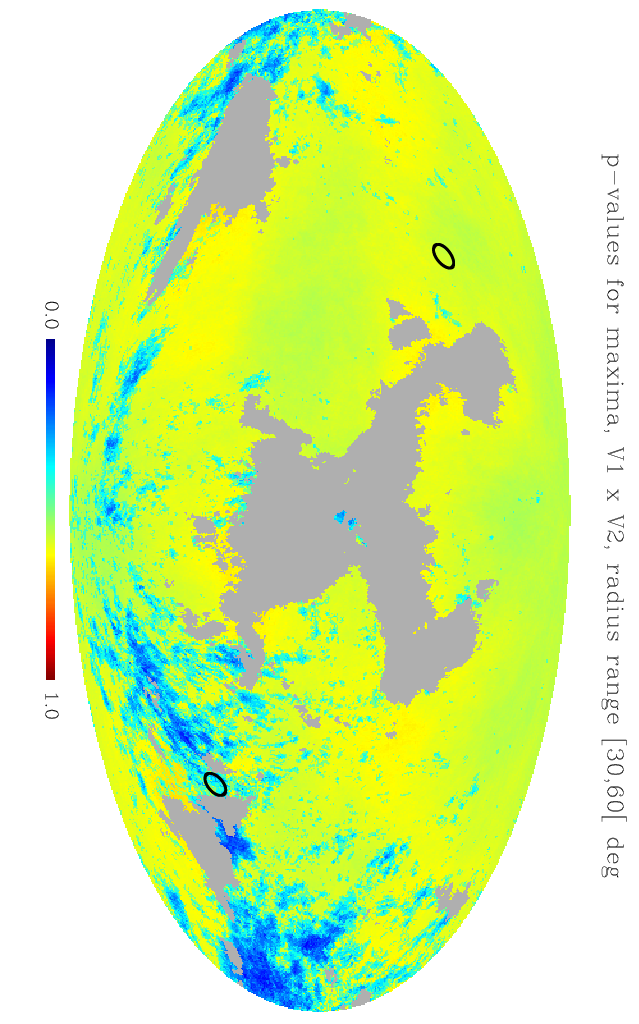,scale=0.18,angle=90}
\epsfig{file=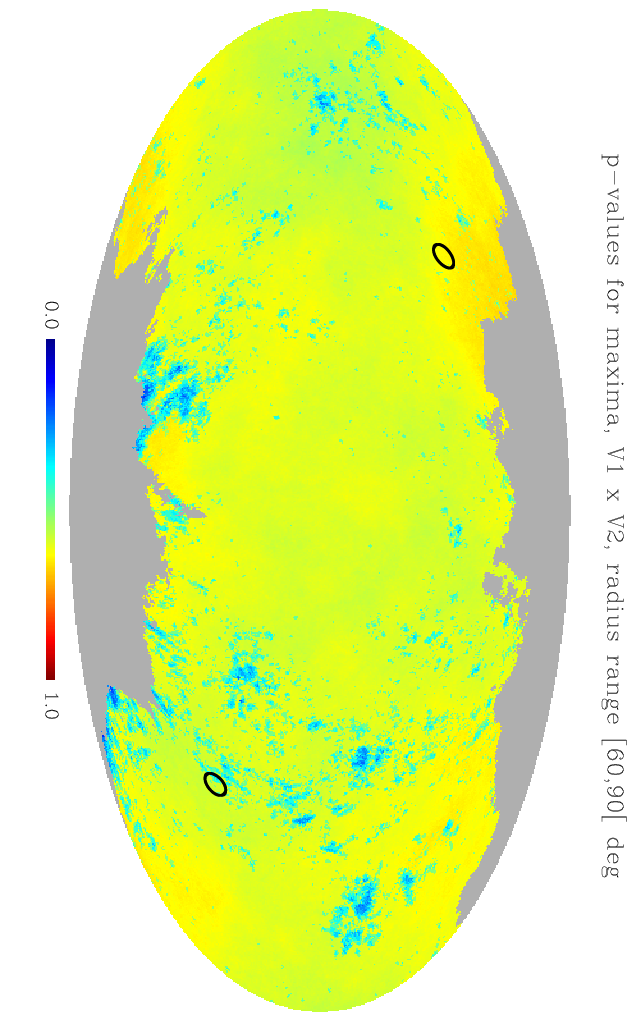,scale=0.18,angle=90}
}
\centerline{
\epsfig{file=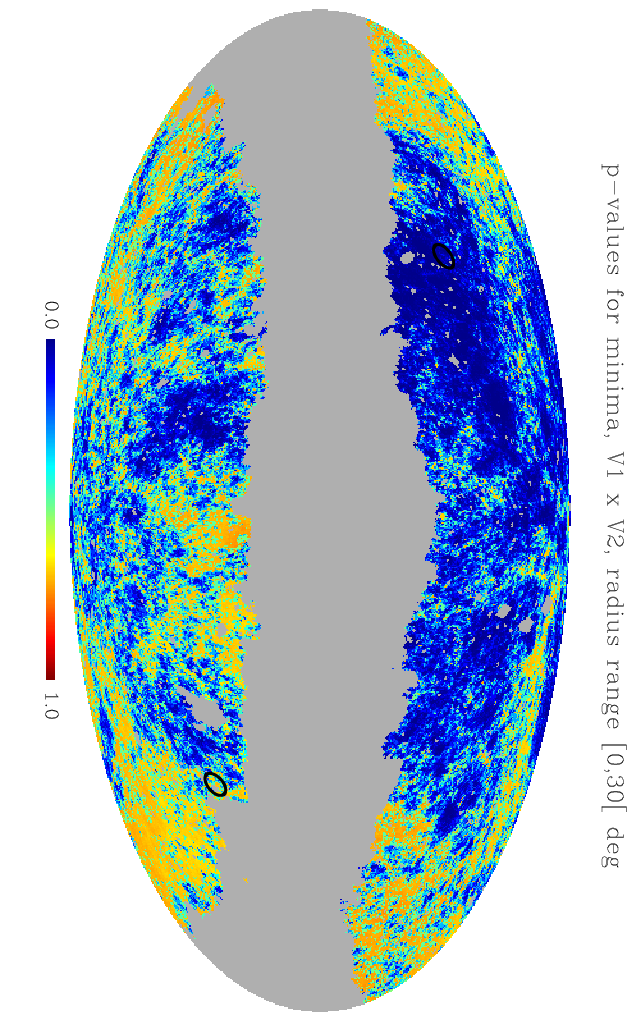,scale=0.18,angle=90}
\epsfig{file=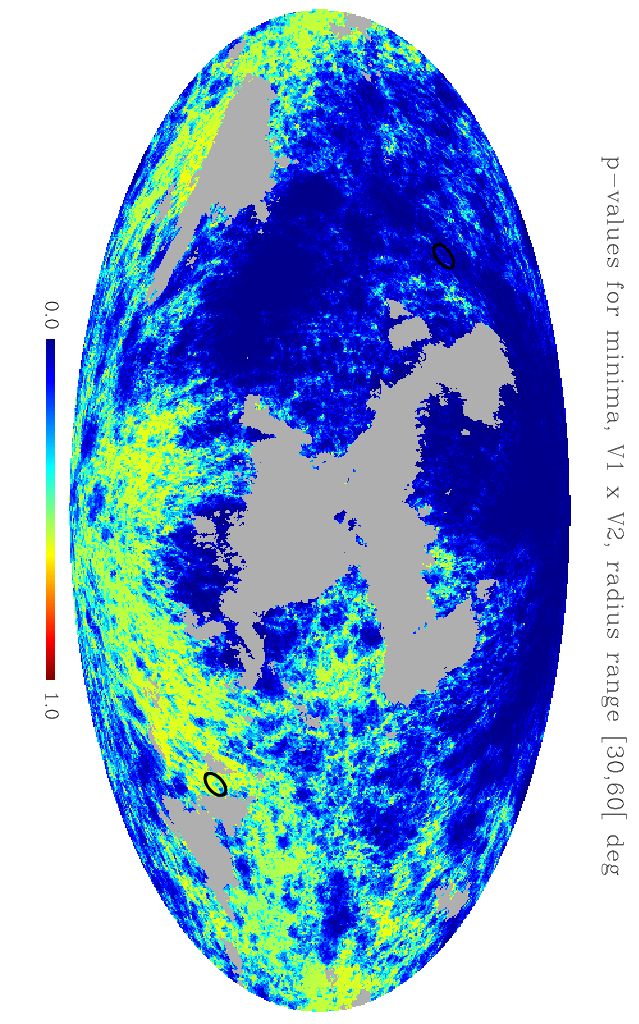,scale=0.18,angle=90}
\epsfig{file=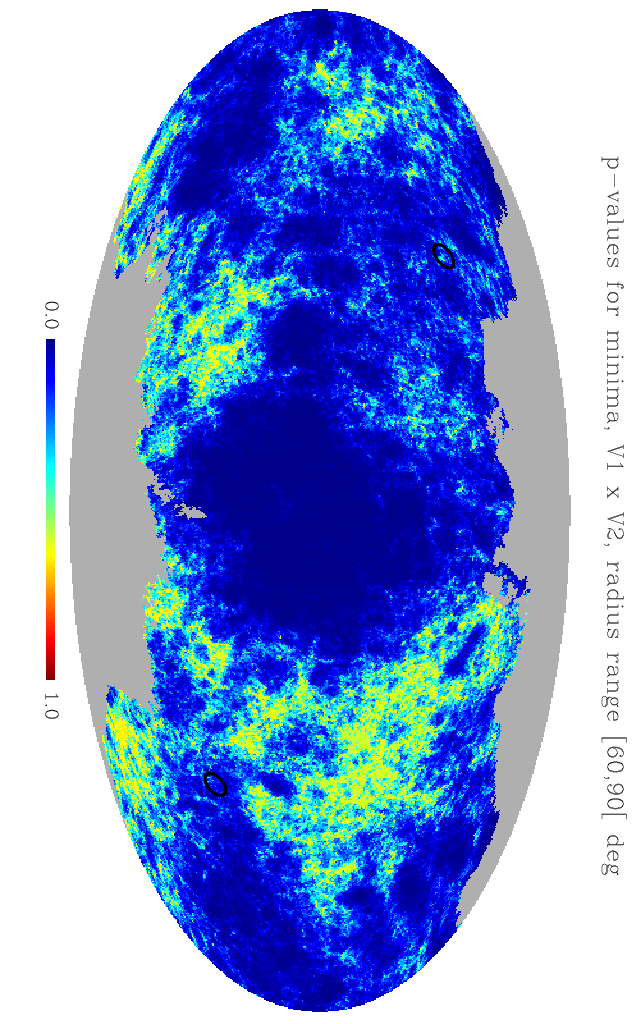,scale=0.18,angle=90}
}

\caption{The p-values map, estimated from 5000 MC simulations,
  for rings with extremely high (upper row) and low (lower row) variances for the three
  intervals of the ring radii: $[0^\circ,30^\circ[$ (left column) ,
  $[30^\circ,60^\circ[$ (middle column) and
  $[60^\circ,90^\circ[$ (right column). The variance map was estimated
  using the V1 and V2 differencing assemblies \emph{WMAP} maps. The black
circles are centred at two points analysed in
Fig.~\ref{fig:variance_profile}. The grey region
 corresponds to pixels for which the p-values were not computed due to
 the  applied mask.} 
\label{fig:pvalues_map_vband}
\end{figure*}

\fig\ref{fig:pvalues_map_vband} shows the probabilities for getting a
larger number of extreme rings centered at a given point for the
V-band maps. Regions with unusually numerous extreme variance rings,
such that none of the 5000 MC simulations show a larger number of
extreme rings, are denoted by the navy blue colour. The colour range
from light blue to red denotes regions with a typical or smaller than
typical number of extreme variance rings for the $\Lambda$CDM model.

In the case of the interval $[0^\circ,30^\circ[$, we can notice that
over a large part of the sky, and especially in the northern Galactic
hemisphere, the number of rings with extremely high variance is
slightly smaller than for most of simulations. However, one can find
also regions, mostly in the southern Ecliptic hemisphere, where the
number of high variance rings is much larger than for the
simulations. As one would expect, these regions are clearly
anti-correlated with the parts of the sky corresponding to rings with
extremely low variance seen in the lower row of the figure
i.e.~regions with a higher number of extremely high variance rings
correspond to regions with lower number of extremely low variance
rings and vice-versa.

One should notice that the point $(l,b)=(105^\circ,37^\circ)$,
corresponding to the center of a set of concentric rings with
systematically smaller variances, lies in one of the region close to
the north Ecliptic pole with many extremely low variance rings. Any
other point in this region shows a similar number of unusually low
variance rings. Such a clustering of rings with low variance is not
consistent with the conformal cyclic cosmology scenario. Since the
positions corresponding to the massive black holes which generate low
variance rings through collisions are distributed randomly on the sky,
the centers of the rings should also be distributed
randomly. Clustering of the rings in relatively large regions rather
indicates effects related to large angular scale features in the CMB
maps.

The anti-correlation between high variance and low variance ring
regions is seen also for the intervals $[30^\circ,60^\circ[$ and
$[60^\circ,90^\circ[$. However, in these cases the p-values for the
number of extremely high variance rings are smaller than for the
interval with radii $[0^\circ,30^\circ[$.This is a consequence of the
stronger correlations between rings of different radius, as the
variance for these intervals is more sensitive to the large angular
scales.  One can also notice fewer regions with small p-values than
for the interval $[0^\circ,30^\circ[$. Conversely, over a large part
of the sky and especially for the interval $[30^\circ,60^\circ[$ and
the northern Ecliptic hemisphere, the number of extremely low variance
rings is significantly larger than for simulations. In the case of the
interval $[60^\circ,90^\circ[$ the low variance rings are centered
mostly in the vicinity of the Galactic center. This disparity in size
of regions with extremely high and low variance rings is related to
the deficit of power at large angular scales in comparison to the
best-fitting $\Lambda$CDM model.

\begin{figure*}

\centerline{
\epsfig{file=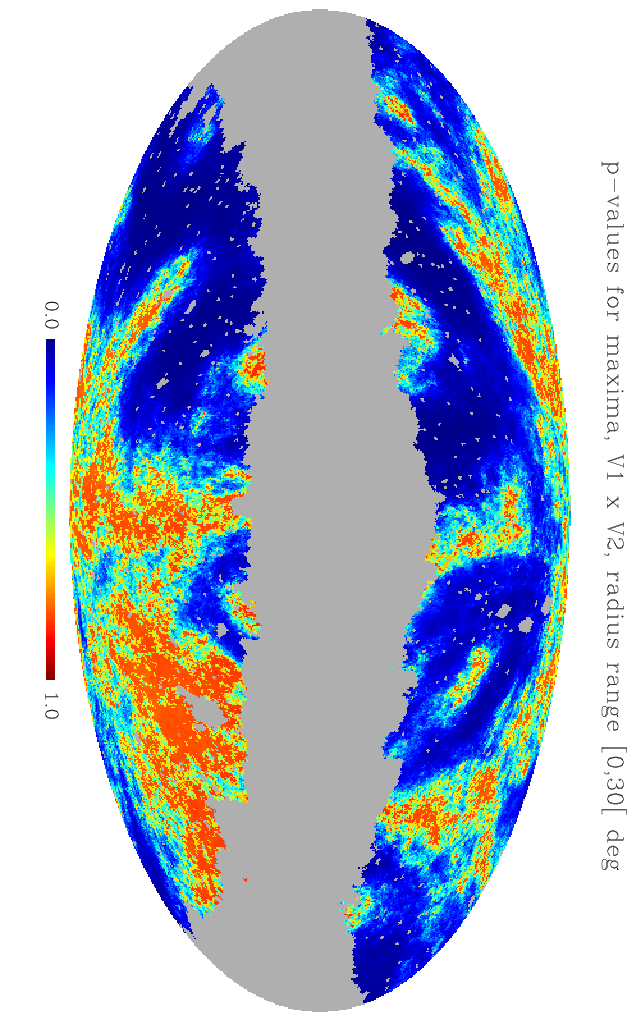,scale=0.18,angle=90}
\epsfig{file=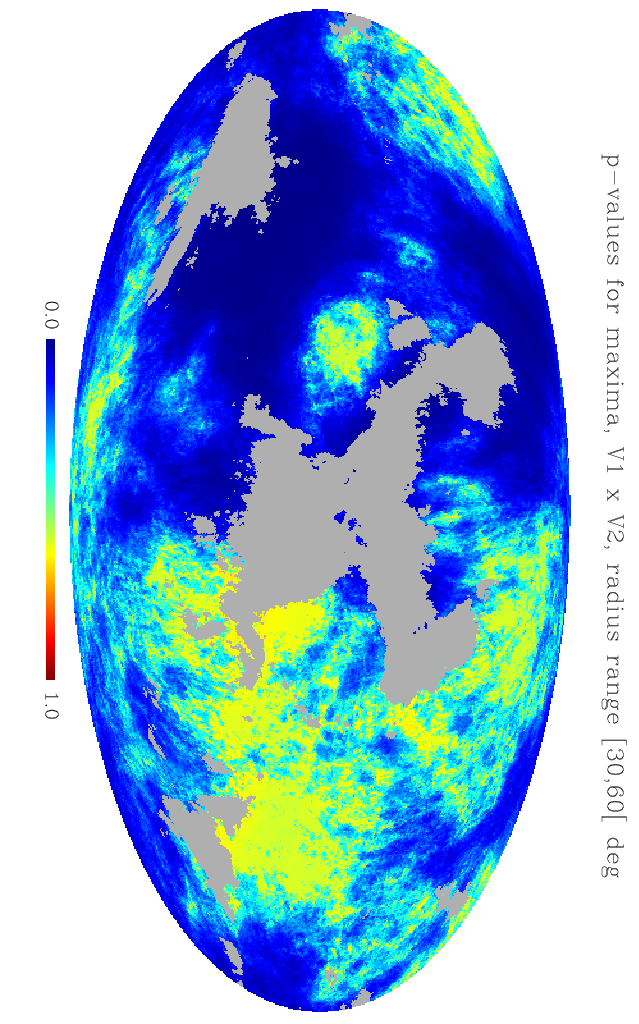,scale=0.18,angle=90}
\epsfig{file=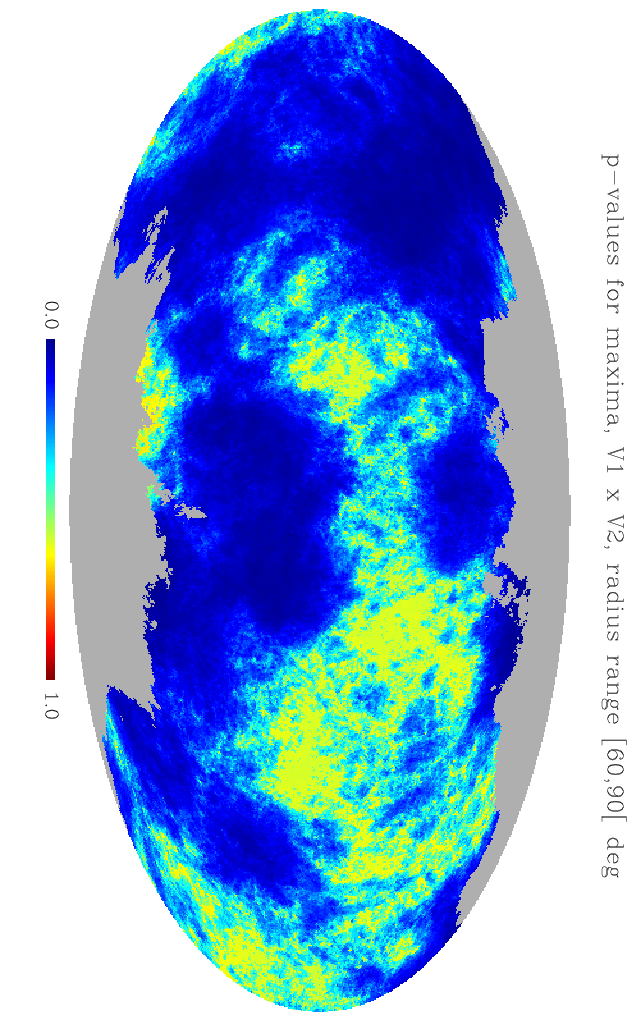,scale=0.18,angle=90}
}
\centerline{
\epsfig{file=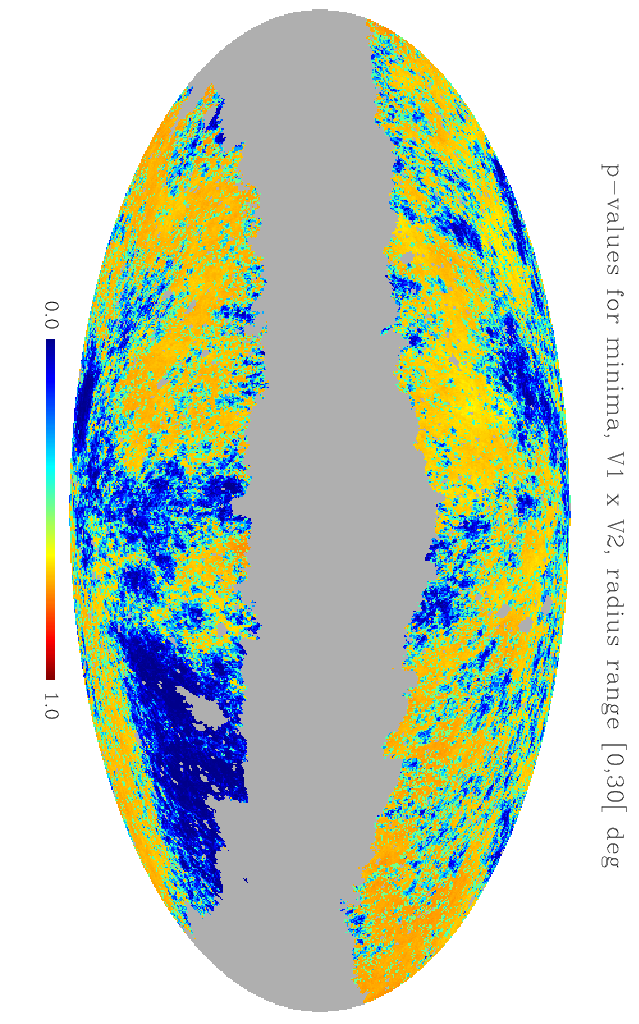,scale=0.18,angle=90}
\epsfig{file=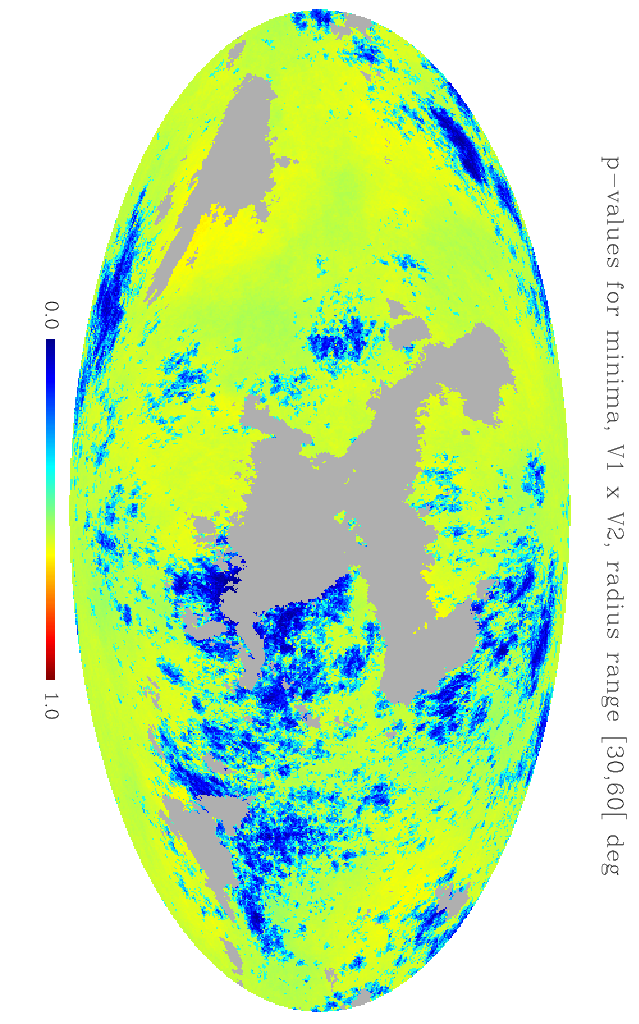,scale=0.18,angle=90}
\epsfig{file=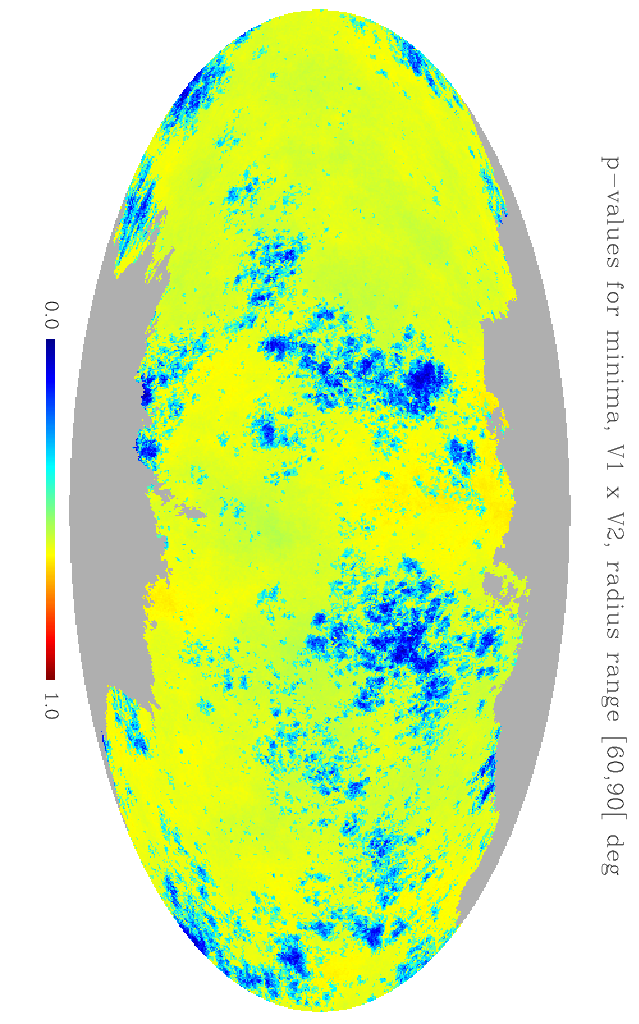,scale=0.18,angle=90}
}

\caption{The same as in Fig.~\ref{fig:pvalues_map_vband} but for
  simulated V1 and V2 DAs.} 
\label{fig:pvalues_sim_map_vband}
\end{figure*}

For comparison we show in \fig\ref{fig:pvalues_sim_map_vband} the
p-value maps for simulated V1 and V2 DA maps with the amplitude of the
quadrupole close to the average predicted for the concordance
$\Lambda$CDM model. As we see, contrary to the \emph{WMAP} data, in
this case regions with small p-value are much larger for extremely
high variance rings and much smaller for extremely low variance
rings. This is seen very well for the radius intervals
$[30^\circ,60^\circ[$ and $[60^\circ,90^\circ[$.  This comparison
confirms again the deficit of power of the \emph{WMAP} maps at large
angular scales. It also shows that relatively large regions with very
small or large probability of getting larger number of extreme rings
are generic. Thus, the p-value maps are rather not useful for
estimation of how significant is this deficit. However, they can be
used to find parts of the sky with a large number of extreme variance
rings that may be interesting for the search for signatures of
alternative models of structure formation such as conformal cyclic
cosmology.

\begin{figure*}

\centerline{
\epsfig{file=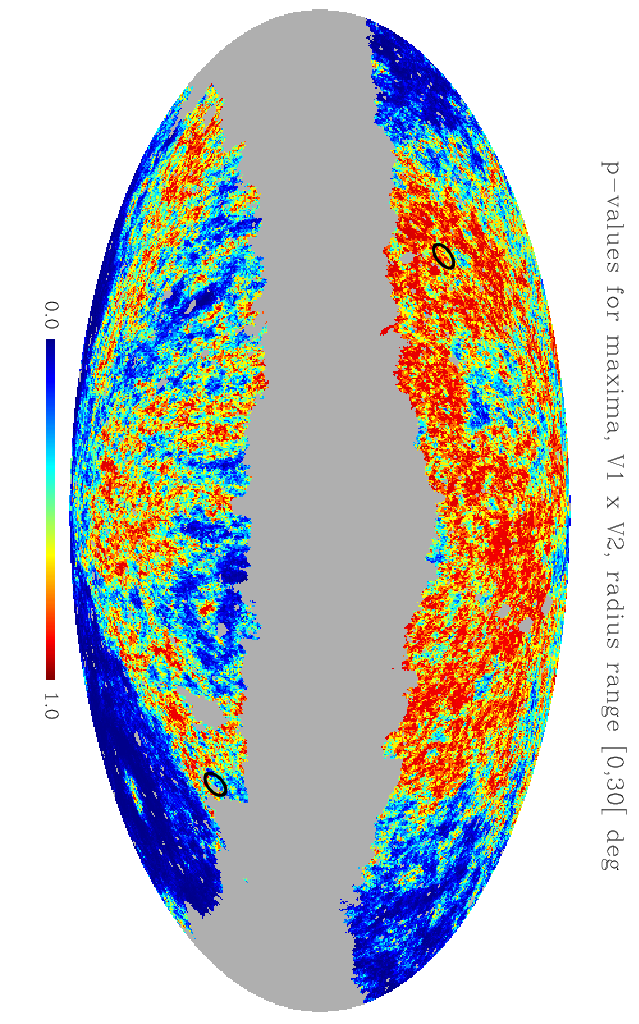,scale=0.18,angle=90}
\epsfig{file=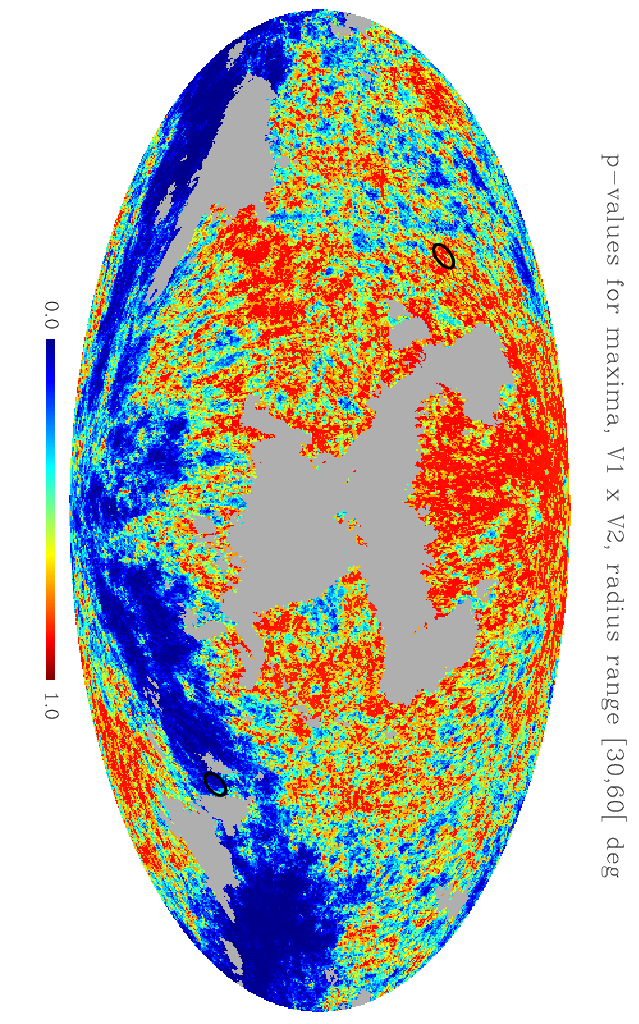,scale=0.18,angle=90}
\epsfig{file=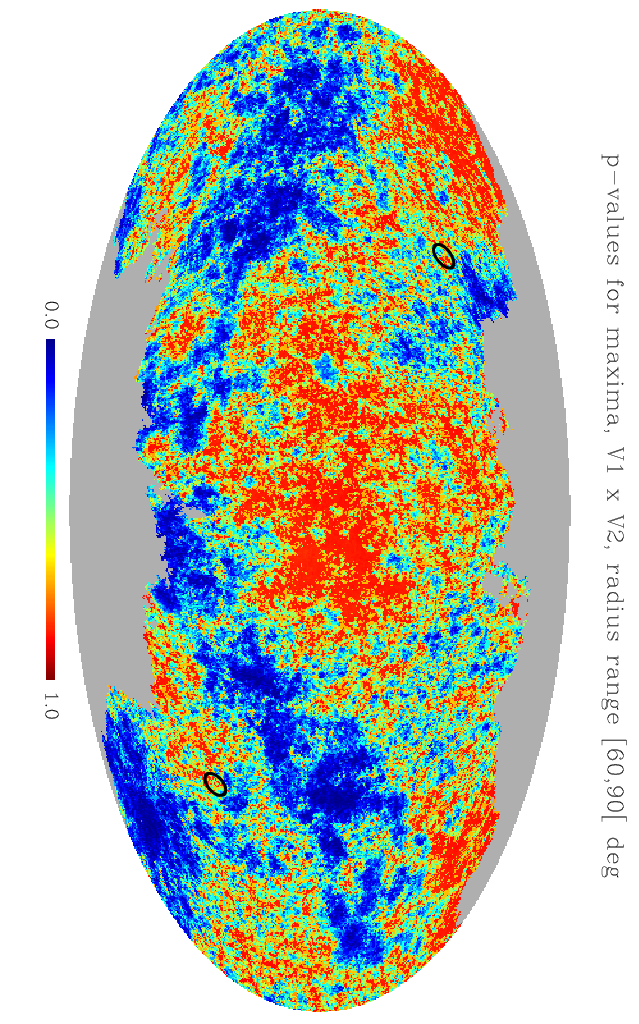,scale=0.18,angle=90}
}
\centerline{
\epsfig{file=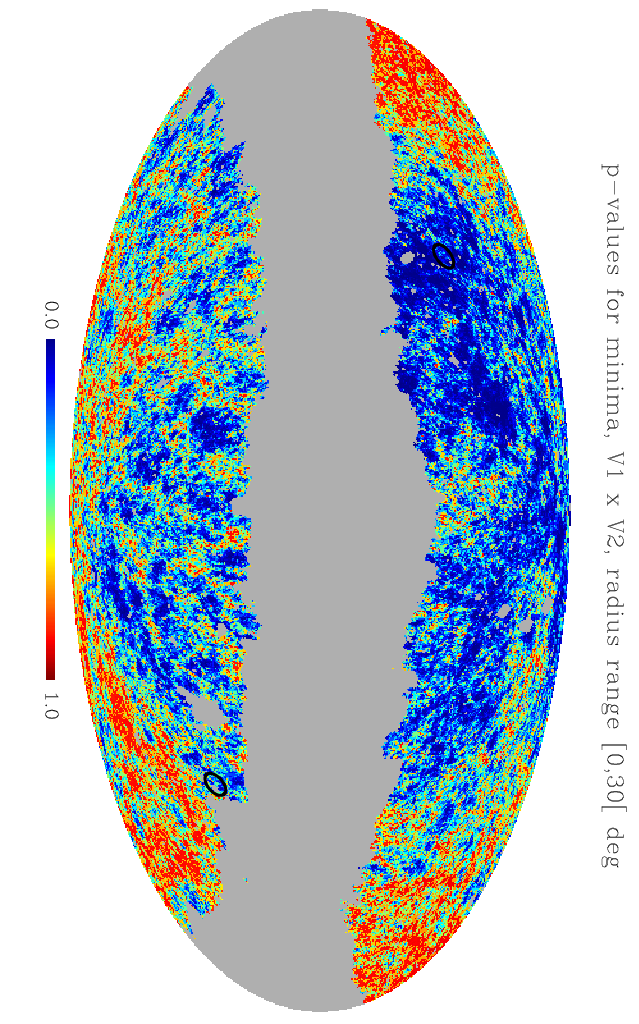,scale=0.18,angle=90}
\epsfig{file=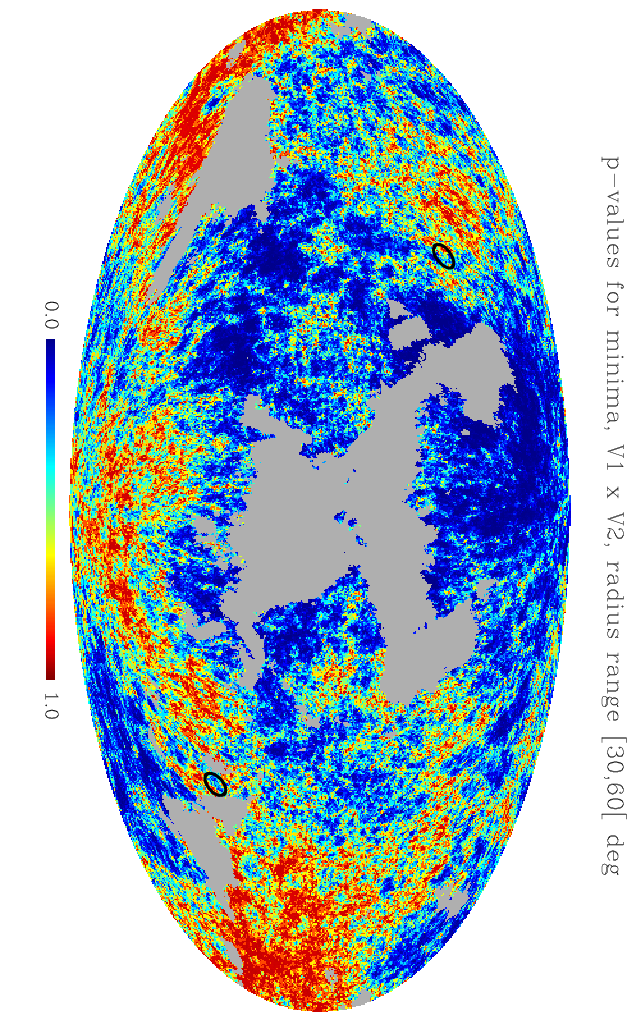,scale=0.18,angle=90}
\epsfig{file=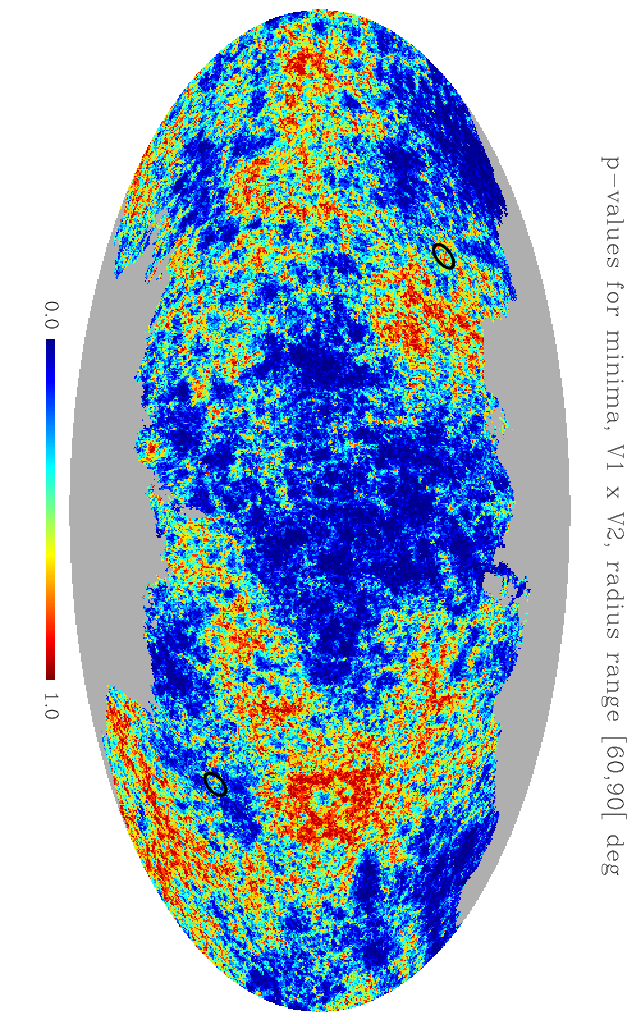,scale=0.18,angle=90}
}

\caption{The same as in Fig.~\ref{fig:pvalues_map_vband} but for
  maps with the best-fitting multipoles in the range $\ell \in [0,7]$ removed.} 
\label{fig:pvalues_map_vband_l7removed}
\end{figure*}

As for the $\chi^2$ map, we repeated estimation of the p-value maps
after removing the best-fitting multipoles in the range $\ell \in [0,7]$
from both simulations and the V-band DAs. The results are shown in
\fig\ref{fig:pvalues_map_vband_l7removed}.  Since the variance for
rings with radii in the interval $[0^\circ,30^\circ[$ is the least
sensitive to the large angular scales, removal of the lowest order
multipoles does not significantly change the p-value maps for this
interval. The most noticeable change is the reduction of the region of
the low variance rings close to the north Ecliptic pole. More
significant changes are noticeable for the other two intervals. Since
the correlations between rings with different radii after removing of
the lowest order multipoles are smaller, the distribution of the
p-values looks more similar to the distribution for the interval
$[0^\circ,30^\circ[$. In the case of the high variance rings, we can also notice extended
regions with lower p-values in the southern Galactic hemisphere. On
the other hand, regions corresponding to the extremely low variance
rings are smaller.

\begin{figure*}

\centerline{
\epsfig{file=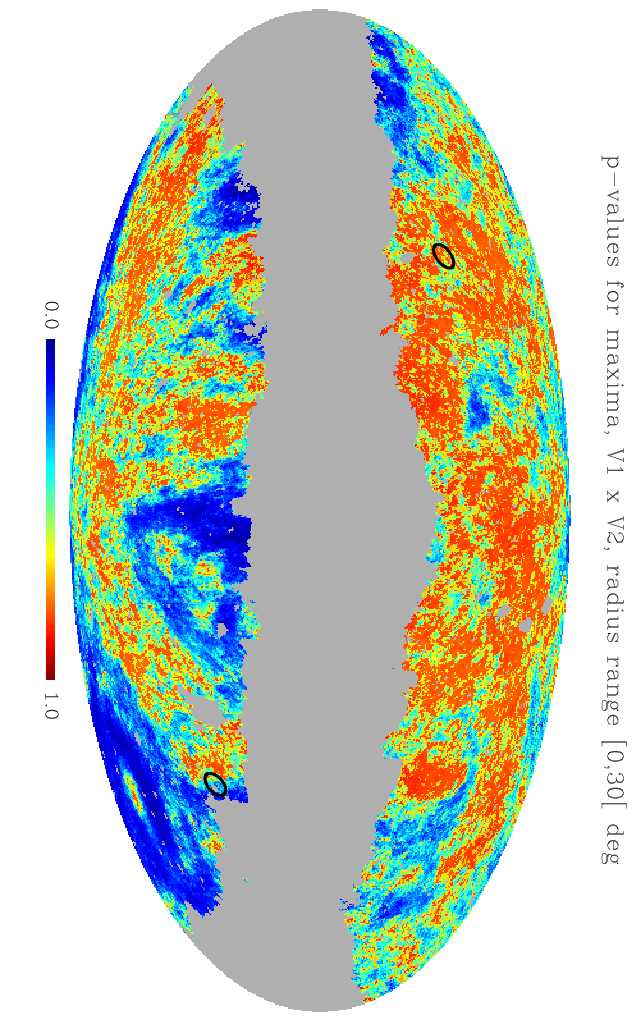,scale=0.18,angle=90}
\epsfig{file=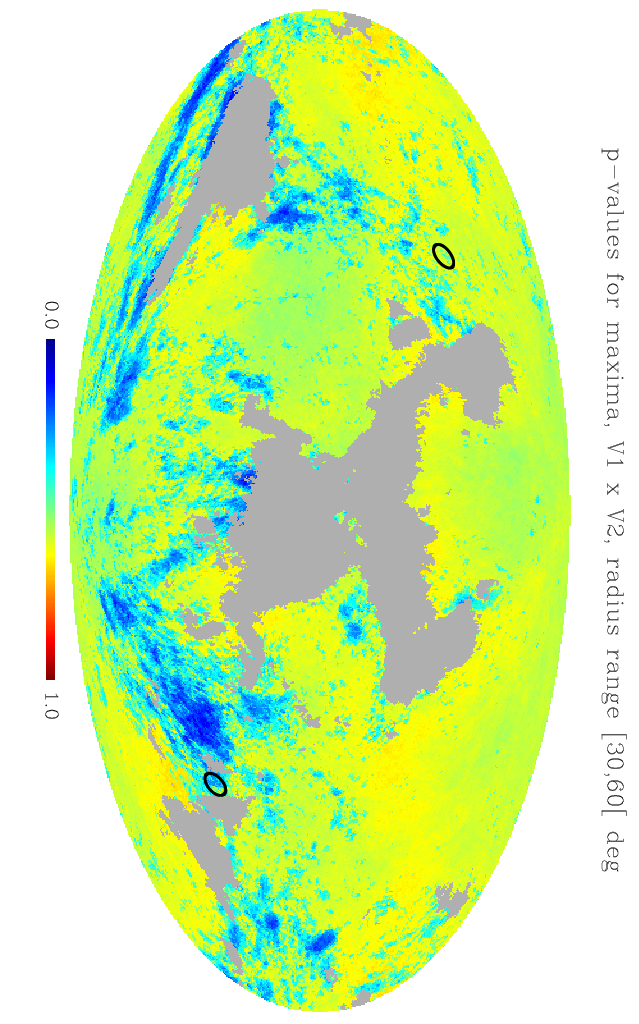,scale=0.18,angle=90}
\epsfig{file=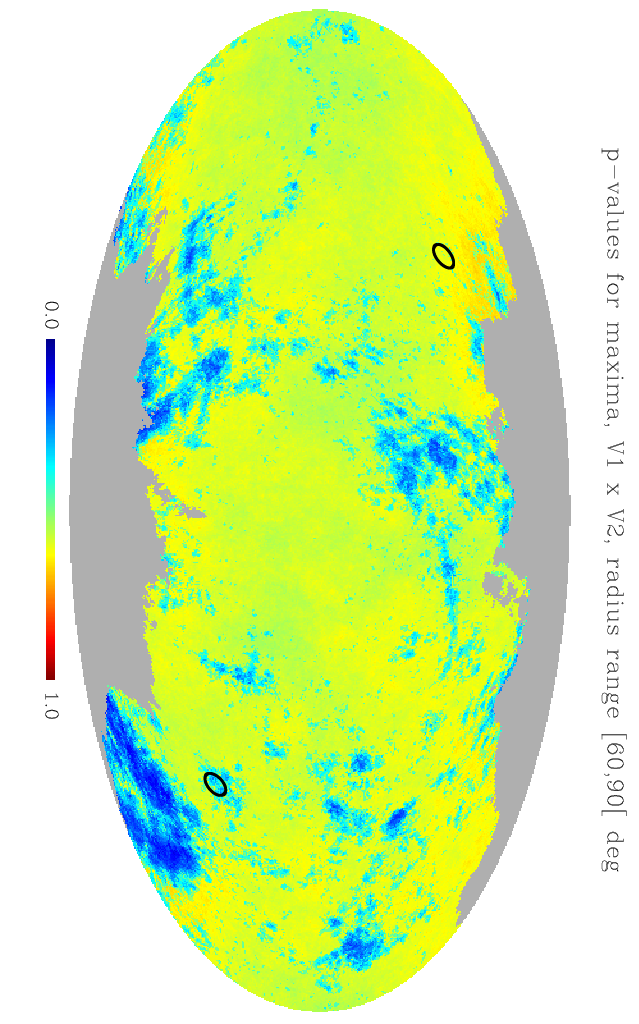,scale=0.18,angle=90}
}
\centerline{
\epsfig{file=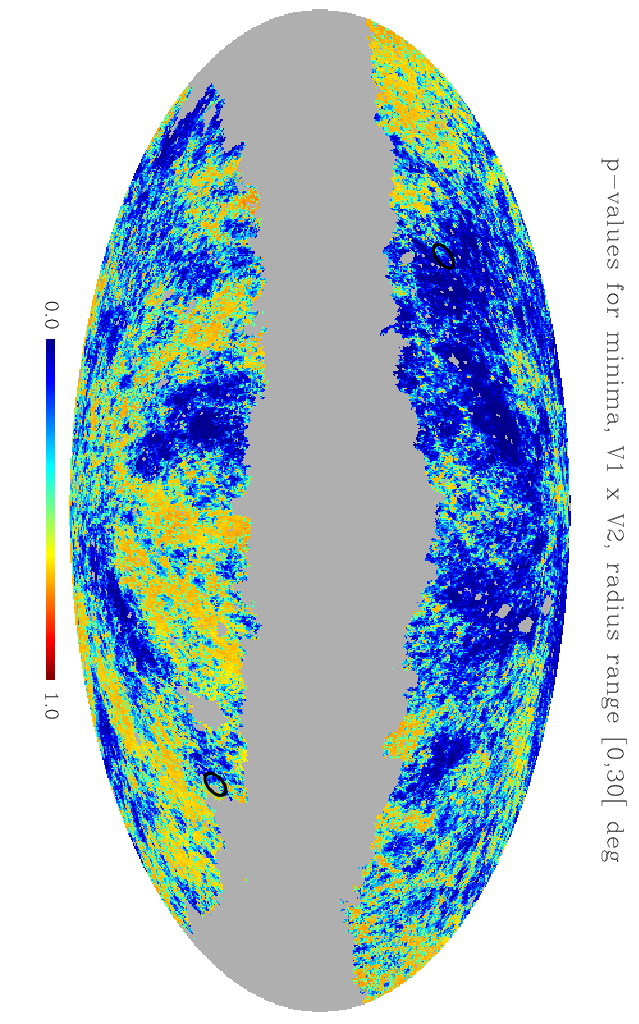,scale=0.18,angle=90}
\epsfig{file=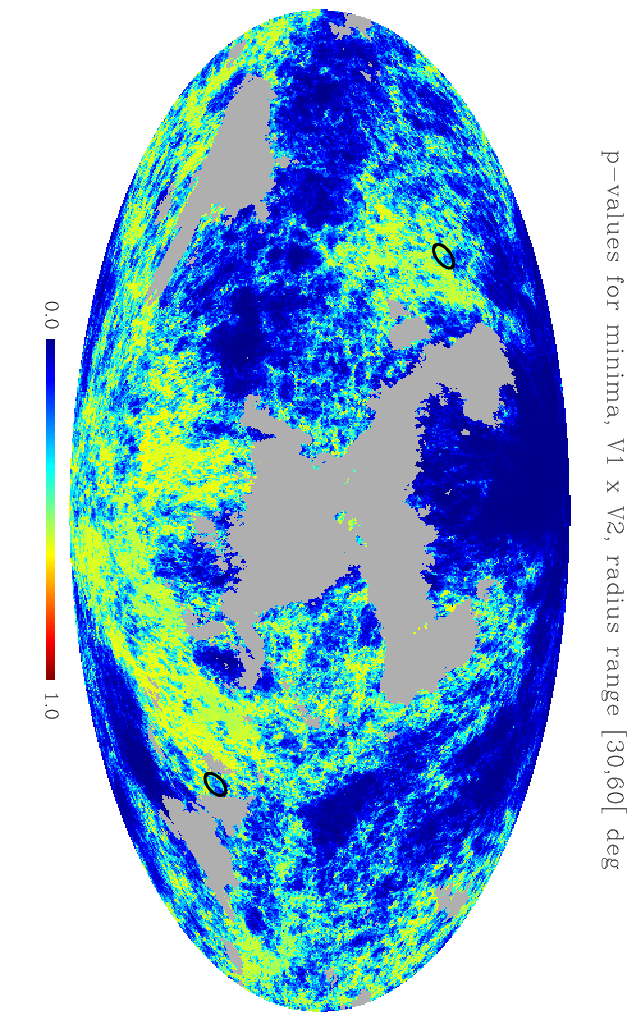,scale=0.18,angle=90}
\epsfig{file=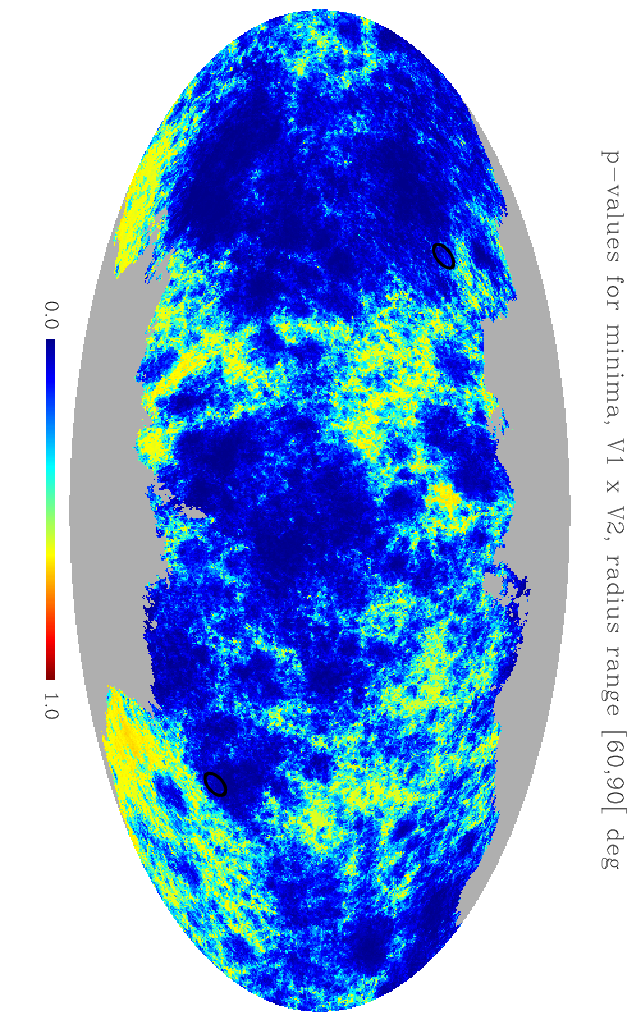,scale=0.18,angle=90}
}

\caption{The same as in Fig.~\ref{fig:pvalues_map_vband} but after rotating the
  best-fitting multipoles in the range $\ell \in [0,7]$
  by  $120^\circ$ around the z-axis. 
}
\label{fig:pvalues_map_rot}
\end{figure*}

Finally, we undertook another test of dependence of the results on the
low-order multipoles. We replaced the best-fitting multipoles in the range
$\ell \in [0,7]$ by the same multipoles rotated around the z-axis by
120$^\circ$. The results for such maps are shown in
\fig\ref{fig:pvalues_map_rot}. After rotation, regions with extremely
low variance rings, especially for the intervals $[30^\circ,60^\circ[$
and $[60^\circ,90^\circ[$, are significantly smaller. However, for the
high variance rings, the changes are less significant.
This test would appear to again confirm that, to a large extent, the
observed extreme variances of the rings are caused by the lowest order
multipoles.

\section{Conclusions}

We have presented a fast method for the computation of variance on
azimuthally symmetric patches of the sky using a convolution
approach.  This has enabled the search for a set of concentric rings
of width of $0.5^\circ$ centred at the nodes of a higher resolution
grid than would be possible for the pixel-based approach used in
previous studies. It has also facilitated the analysis of rings over
the full range of radii from 0 to 90 degrees.  The technique was
applied to the 7-year \emph{WMAP} data to search for rings with
unusually low or high variance.

Analysis of the variance of rings centred at two points,
$(l,b)=(105^\circ,37^\circ)$ and $(l,b)=(252^\circ,-31^\circ)$, which
have recently drawn special attention in the context of the conformal
cyclic cosmology scenario, showed that for the former with ring radius
larger than about $10^\circ$ the variance is systematically smaller
than the average from MC simulations. However, after replacing the
$\ell \in [0,7]$ multipoles in the simulations with the best-fitting values
from the foreground corrected V-band data masked with the KQ75y7 cut,
the agreement between simulations and the observed variances is very
good. Therefore, we can conclude that observed deficit of power in
rings with large radii is related to the anomalous distribution of the
lowest order multipoles of the \emph{WMAP} map. Moreover, the p-value
maps for the number of extremely low variance rings show that one can
easily find points with as many anomalously low variance rings as for
the point $(l,b)=(105^\circ,37^\circ)$. Thus, there is nothing unusual in the properties
of the rings centred at this point given the observed large scale
anomalies. These results are not consistent 
with predictions of the conformal cyclic cosmology scenario and they
can not be used to support it. Whilst this model predicts randomly
distributed sets of concentric rings within which few have lower
variance, we observe sets of rings with systematically lower variance
for larger ring radii that are clustered in a few relatively large
regions of the sky

A strong dependence of the variance in rings, especially with large
radius, on the lowest order multipoles is also seen for other
statistics used in our studies. The $\chi^2$ map corresponding to the
radius intervals $[30^\circ,60^\circ[$ and $[60^\circ,90^\circ[$ takes
smaller values than simulations for the best-fitting $\Lambda$CDM model at
a significance level around 95\%. Furthermore, in a large part of the
northern Ecliptic and Galactic hemispheres, the probability of getting
a larger number of extremely low variance rings is very low, while
extremely high variance regions are much smaller and localised mostly
in the southern Ecliptic hemisphere. All of these anomalies disappear
after removing from the maps the best-fitting multipoles in the range
$\ell \in [0,7]$. The disparity in distribution of the low and high
variance rings is also weaker after rotation of the lowest order
multipoles around the Galactic poles. Thus, the anomalous variances in
rings can be traced to the presence of anomalous low-order multipoles.

However, as we have verified, the removal of the quadrupole and octopole is
insufficient to completely eliminate the anomalies. Consequently,
we have found that they are also related to higher order multipoles,
up to $\ell=7$. This conclusion is consistent with the results of
previous studies on the large angular scale anomalies observed in the
\emph{WMAP} data \citep{hou:2009,hou:2010}. Studies of the physical origin of these
anomalies is beyond the scope of this paper. However, they do not seem
to be related to residuals of the Galactic thermal dust emission since
the results for the W-band map are consistent with the results for the
V-band map.

The fast method of computation of variance presented here can be used for
any azimuthally symmetric patches such as discs with an arbitrary radial
weighting function. It can also be easily extended to computations of
higher order moments enabling various statistical tests of the data on
the sphere with high angular resolution. Furthermore, with some
performance penalty, the method can be 
extended to the calculation of moments on non-azimuthally symmetric patches
using the more general convolution approach proposed by
\citet{wandelt:2001}.

\section*{Acknowledgments}
We acknowledge use of CAMB \citep{camb} and the HEALPix software
\citep{gorski:2005} analysis package for deriving the results in this
paper. We acknowledge the use of the Legacy Archive for
Microwave Background Data Analysis (LAMBDA). Support for LAMBDA is
provided by the NASA Office of Space Science. This research was
supported by the Agence Nationale de la Recherche (ANR-08-CEXC-0002-01).

%------------------------------------

\begin{appendix}

\section{Estimation of the low-order multipoles using a direct
  inversion method} \label{sec:appendix}

The low-order multipoles can be reconstructed from the cut sky
  map using the following relation between the cut sky spherical harmonic
  coefficients $\widetilde{a}_{\ell m}$ and the full sky
  coefficients $a_{\ell m}$ :
\begin{equation} \label{eqn:pseudoalms}
\widetilde{a}_{\ell m} = \sum_{\ell' m'} K_{\ell m,\, \ell' m'}\ a_{\ell' m'} ,
\end{equation}
where
\begin{equation}
K_{\ell m,\, \ell' m'} \equiv \int_{\rm{cut\ sky}} Y_{\ell m}^\ast (\hat{\mathbf{n}}) 
Y_{\ell' m'}^{}(\hat{\mathbf{n}})\ d \Omega_{\hat{\mathbf{n}}} \ ,
\end{equation} 
is the spherical harmonics coupling matrix for the cut sky. We assume
  that noise is negligible for the low-order multipoles thus it is not
  included in the above relation.
 
To estimate the full sky multipole coefficients $\hat{a}_{\ell m}$ one
just needs to invert the coupling matrix $K_{\ell m,\, \ell' m'}$ :
\begin{equation} \label{eqn:direct_inversion}
\hat{a}_{\ell m} = \sum_{\ell' m'} K_{\ell m,\, \ell' m'}^{-1}\
\widetilde{a}_{\ell' m'} \ .
\end{equation}
In general, the sky cut causes the coupling matrix $K_{\ell m,\, \ell'
  m'}$ to be singular and it is impossible to reconstruct all of the modes from 
the $\widetilde{a}_{\ell m}$. However, for low-order multipoles and small
sky cuts, a good approximation is to simply truncate
the coupling matrix at some multipole $\ell$ and invert the non-singular
truncated matrix.

This estimator provides the full sky $a_{\ell m}$
coefficients that are the best fits to the cut sky map
in the sense of minimising the function
\begin{equation} \label{eqn:chisq_alms}
\chi^2 (a_{\ell'' m''}) = \sum_{\ell m} \left( \widetilde{a}_{\ell m} -
  \sum_{\ell' m'} K_{\ell m,\, \ell' m'}\ a_{\ell' m'} \right)^2 \ .
\end{equation}
It is worth noticing that this method is a straightforward
  generalization for higher order multipoles of the approach employed
  in the {\it remove\_dipole} routine from the \textsc{healpix} package. We refer to
  this method as direct inversion.

\end{appendix}

\end{document}